\documentclass[prb,aps,english,twocolumn,notitlepage,superscriptaddress]{revtex4-2}
\usepackage{graphicx}
\usepackage{dcolumn}
\usepackage{rotating}
\usepackage{lipsum}
\usepackage{xcolor}
\usepackage{array}
\usepackage{bm}
\usepackage[T1]{fontenc}  
\usepackage[latin1]{inputenc}
\usepackage{multirow,amsmath,amssymb}
 \usepackage{babel} 
 \usepackage{txfonts}
 \usepackage{epsfig,epsf,psfrag} 
\usepackage{graphicx} 
\usepackage{pslatex}
\usepackage{epic,eepic} 
\usepackage{color,pstcol}
\usepackage{pstricks} 
\usepackage{fancyhdr} 
\usepackage{tabularx}
\usepackage{physics}
\usepackage{url}
\usepackage{listings}
\usepackage{color} 
\usepackage{caption}
\usepackage{balance}
\usepackage{subcaption}

\begin{document}
\title{Majorana fermion induced power-law scaling in the violations of the Wiedemann-Franz law}

\author{Sachiraj Mishra}
\email{sachiraj.mishra@niser.ac.in}
\affiliation{%
School of Physical Sciences, National Institute of Science Education \& Research, Jatani 752050, India\\ Homi Bhabha National Institute, Training School Complex, Anushaktinagar, Mumbai 400094, India
}%
\author{Ritesh Das}
\email{ritesh.das@niser.ac.in}
\affiliation{%
School of Physical Sciences, National Institute of Science Education \& Research, Jatani 752050, India\\ Homi Bhabha National Institute, Training School Complex, Anushaktinagar, Mumbai 400094, India
}%
\author{Colin Benjamin}%
\email{colin.nano@gmail.com}
\affiliation{%
School of Physical Sciences, National Institute of Science Education \& Research, Jatani 752050, India\\ Homi Bhabha National Institute, Training School Complex, Anushaktinagar, Mumbai 400094, India
}%
\begin{abstract}
Violation of the Wiedemann-Franz law in a 2D topological insulator due to Majorana bound states  is studied via the Lorenz ratio in the single-particle picture. We study the scaling of the Lorenz ratio in the presence and absence of Majorana bound states with inelastic scattering modeled using a Buttiker voltage-temperature probe. We compare our results with that seen in a quantum dot junction in the Luttinger liquid picture operating in the topological Kondo regime. {We explore the scaling of the Lorentz ratio in our setup when either phase and momentum relaxation or phase relaxation is present. This scaling differs from that predicted by the Luttinger liquid picture for both uncoupled and coupled Majorana cases.}.
\end{abstract}
\maketitle
\section{Introduction}
Majorana fermions are a particular class of fermions with the unique property of being their antiparticles. Majorana fermions have intrinsic topological protection from disorder. This property renders Majorana fermions immune to decoherence, making Majorana fermions an ideal candidate for quantum computation. {These particles manifest as a zero-energy mode in the $p_x + i p_y$ wave superconductor as initially explained by Kitaev \cite{kitaev2001unpaired}. Moreover, their presence has been anticipated in 2D topological insulators interfaced with $s$-wave superconductors and separated by a ferromagnet \cite{PhysRevLett.100.096407, PhysRevLett.101.120403}.  
Numerous promising theoretical proposals have been suggested for the detection of Majorana bound states (MBS). By examining the conductance, distinctions between topological and trivial superconductors can be made, aiding in MBS detection \cite{PhysRevLett.98.237002, PhysRevLett.103.237001}. Hanbury-Brown and Twiss (HBT) non-local shot noise-like correlations \cite{PhysRevB.106.125402}, along with non-local conductance \cite{PhysRevB.81.085101}, have proven effective in distinguishing between topological and trivial superconductors, offering a reliable method for detecting MBS. Similarly, a Josephson junction modeled by superconductor-topological insulator-superconductor can exhibit a $4 \pi$ periodic current-phase relationship instead of $2 \pi$, attributed to MBS \cite{PhysRevB.79.161408}}. Further, in semiconductor-superconductor heterostructures \cite{Lutchyn2018}, MBS are shown to occur. There have been numerous attempts to experimentally detect Majorana fermions \cite{retraction2, gazibegovic2017retracted, yin2015observation, Nayak_2021}, but irrefutable evidence for their existence remains elusive. {In \cite{retraction2}, Majorana zero modes were initially reported based on the observation of a zero-bias conductance peak (ZBCP) quantized at $2e^2/h$ in a semiconductor nanowire coupled with an $s$-wave superconductor. However, this claim was later retracted. Similarly, Ref.~\cite{gazibegovic2017retracted} claimed the detection of Majorana zero modes in a hybrid semiconductor-superconductor interface by studying the Aharonov-Bohm effect and weak antilocalization, but this claim was also retracted. Other experimental works \cite{yin2015observation, Nayak_2021} have also focused on ZBCPs as potential indicators of Majorana fermions, although these peaks alone may not definitively confirm the presence of Majorana bound states (MBS). In a recent study \cite{uday2024induced}, a signature of crossed Andreev reflection was observed in a magnetic topological insulator that hosts the quantum anomalous Hall effect when in proximity to an $s$-wave superconductor such as Niobium. This signature of crossed Andreev reflection holds considerable significance for ongoing studies on Majorana bound states. There are some theoretical proposals on the connection between Majorana fermions and quantum metrology too \cite{carollo2020geometry}.}

 There have been proposals to detect Majorana fermions using violations in Wiedemann-Franz law \cite{buccheri} by studying the scaling of the Lorenz ratio. {The Wiedemann-Franz (WF) law states that the ratio of effective thermal conductance to effective electrical conductance is proportional to the temperature of the system.} In our case, the system is an Aharonov-Bohm ring composed of a topological insulator. We define this ratio as the Lorentz ratio. Ref. \cite{buccheri} studies the violation of Wiedemann-Franz law in a quantum dot junction in the topological Kondo regime \cite{kondo.109.156803} hosting localized Majorana bound states (MBS). WF law states that the ratio of the electrical conductance to the thermal conductance is inversely proportional to the temperature. The constant of proportionality, also known as the Lorenz ratio, is a constant for all conductors. Majorana fermions are also known to break particle-hole symmetry (PHS) and induce violations of Wiedemann-Franz (WF) law \cite{buccheri}. In the presence of MBS, the Lorenz ratio has been shown to scale inversely with respect to the Luttinger parameter \cite{buccheri}. The setup in \cite{buccheri} is considered in the many body regime and shows that the Lorenz ratio shows power-law decay as a function of the Luttinger parameter in the setup.

In this paper, we discuss the violation of WF law and the scaling of the Lorenz ratio in the presence of Majorana fermions using a Buttiker voltage temperature {(BVT)} probe \cite{buttikerboth, PhysRevB.33.3020, dvira, kilgour2015charge, kilgour2015tunneling}, which induces inelastic scattering. We consider two kinds of inelastic scattering in our setup: phase and momentum relaxation \cite{buttikerboth, heikkila2013physics} and with only phase relaxation \cite{PhysRevB.33.3020}. We study the scaling of the Lorenz ratio with respect to the strength of inelastic scattering and compare the scaling of the Lorenz ratio with the Luttinger parameter in the many-body setup considered in \cite{buccheri}. In our setup, we study three cases: (i) when MBS are absent, (ii) when MBS are present and uncoupled, and (iii) when MBS are coupled. In a single-electron setup, the Lorenz ratio scales differently with inelastic scattering when both phase and momentum relaxation occur and when only phase relaxation occurs. For individual MBS, WF law is not violated without inelastic scattering. 

The rest of the paper is organized as follows: In section II, we describe the motivation and importance of this work. Section III describes the scattering and transmission in our setup using Landauer-Buttiker scattering formalism \cite{PhysRevB.53.16390,lambert1993multi}. In section IV, we first calculate the thermoelectric coefficients like the Seebeck, Peltier, and thermal conductance in our setup using the Onsager relations. We then define the Lorenz ratio and introduce inelastic scattering via a {BVT} probe \cite{dvira}. In section V, we present an analysis of our results and compare the scaling of the Lorenz ratio with the strength of inelastic scattering and compare it with the scaling seen in Ref. \cite{buccheri}. We end with the conclusions in section VI.

\section{Motivation}

{The primary motivation of our work is indeed to detect Majorana-bound states through the observation of a power-law scaling in the violation of the Wiedemann-Franz law. However, our study also aims to explore whether this power-law scaling, previously observed in the Luttinger liquid regime, is also seen using a phenomenological model of {BVT} probe with edge mode transport in Landaeur-Buttiker scattering theory. This investigation is crucial because the many-body effects that are fundamental in the Luttinger liquid are phenomenologically incorporated into the scattering framework via the {BVT} probe. Our results demonstrate that while the nature of the violation of the Wiedemann-Franz law with respect to the inelastic scattering parameter in a single-particle picture remains similar between the two regimes, the power-law scaling differs. Notably, this work represents the only study where a violation of the Wiedemann-Franz law adheres to a specific scaling law within the Landauer-Buttiker scattering framework, offering a detailed comparison with the Luttinger liquid regime \cite{buccheri}.}

Ref \cite{buccheri} studies a quantum dot junction capable of hosting an even number of Majorana fermions to study the setup in the topological Kondo regime \cite{kondo.109.156803}. In special cases, the electrons in the quantum dot can couple to two-fold degenerate states. The regular Kondo effect \cite{kondokodesuga} is a consequence of the coupling of mobile electrons in a confined region to spin-degenerate states. Further, an even number of Majorana fermions can couple non-locally, giving rise to two-fold degenerate states that can couple to electrons in the quantum dot \cite{kondo.109.156803}. It is known as the topological Kondo effect \cite{kondo.109.156803}. The setup is studied in the Luttinger liquid model, which uses many-body formalism and considers inelastic scattering due to electron-electron interaction. The electron-electron interaction is parameterized by the Luttinger parameter $(g)$, with $g = 1$ corresponding to the absence of interaction, $g < 1$ corresponding to attractive interaction, and $g > 1$ corresponding to repulsive interaction. The authors of Ref. \cite{buccheri} show that the Lorenz ratio $(W)$ depends on the Luttinger parameter as $W(g) =
\frac{2}{3g}$. When the setup in Ref. \cite{buccheri} is in the topological Kondo regime, the Majorana-induced boundary conditions and scattering via a splitting junction leads to the "splitting" of a charged particle into a transmitted particle of charge $\frac{2e}{3}$ and a backscattered hole of charge $\frac{e}{3}$. The coefficient $\frac{2}{3}$ is the unique and universal signature of Majorana bound states in a quantum dot junction operating in the topological regime according to Ref. \cite{buccheri}.
\par
Our setup introduces inelastic scattering phenomenologically using Buttiker voltage probe \cite{PhysRevB.33.3020, buttikerboth}. The Buttiker probe is an additional probe that induces inelastic scattering in the setup such that the total charge and heat current flowing into the Buttiker probe is zero. Unlike Luttinger liquid theory, the Buttiker voltage probe is based on single-particle scattering theory, making it more adaptable to setups studied using scattering theory. Further, using the Buttiker voltage probe, we can study inelastic scattering with phase and momentum relaxation \cite{buttikerboth} and inelastic scattering with only phase relaxation \cite{PhysRevB.33.3020}. In our work, we look at both electric conductance and thermal conductance. Therefore, we modify the Buttiker voltage probe into a {BVT} probe, as was also done recently in \cite{dvira}. The comparison between the Luttinger parameter and inelastic scattering due to the {BVT} probe \cite{dvira} has not been studied before. We study the scaling of the Lorenz ratio (W) with respect to the coupling strength of the {BVT} probe in our setup. A {BVT} probe is added to the setup, such that the total charge and heat currents going into the probe vanish but induce inelastic scattering via both phase and momentum relaxation \cite{buttikerboth} or only phase relaxation \cite{PhysRevB.33.3020}. Any electron/hole entering the Buttiker probe loses its phase memory and is reinjected with a completely different phase, leading to phase relaxation \cite{PhysRevB.33.3020, buttikerboth}. For the case of both phase and momentum relaxation \cite{buttikerboth}, the electrons injected into the probe from the setup are reinjected with equal probabilities of going toward the left or the right. Thus, the reinjection also causes the phase and the initial momentum to be lost. To preserve the momentum while inducing phase relaxation, one can use a pair of unidirectional probes such that when the electrons initially traveling to the left or the right are injected into the Buttiker probe, they are reinjected back with the same momentum, see Ref. \cite{PhysRevB.33.3020}. We seek to probe MBS using two different kinds of {BVT} probes, note the difference in these results, and compare our results with the results seen in Ref. \cite{buccheri}, which is studied in the many body picture. {Generally, in mesoscopic devices, noise and environmental effects \cite{PhysRevB.110.045429} are modeled phenomenologically via the {BVT} probe in the setup with both phase and momentum relaxation and phase relaxation.} This distinguishing behavior of MBS as a function of inelastic scattering in {BVT} probe \cite{dvira} is revealed through the Lorenz ratio. The signatures distinguish MBS's existence and nature (coupled or individual).

\section{Description of the Model}
\subsection{Hamiltonian}
In Fig. 1, we show our proposed model. Our proposed Majorana Aharonov-Bohm interferometer (ABI) is based on helical edge modes generated via the quantum spin Hall effect in topological insulators (TIs). We mold a 2D TI into an Aharonov-Bohm ring wherein spin-orbit coupling generates protected 1D edge modes. The ring is pierced by an Aharonov-Bohm flux $\phi$. The Dirac equation for electrons and holes in the ring is given as
\begin{equation} \label{eq1}
[v p \tau_{z} \sigma_{z} + (- E_{F} + \frac{eA}{\hbar c})\tau_{z}]\psi = E\psi\,.
\end{equation}
$\psi = (\psi_{e \uparrow}, \psi_{e \downarrow},\psi_{h \downarrow},\psi_{h \uparrow})^{T}$ is a four-component spinor, $p = -i \hbar \frac{\partial}{\partial x}$ is the momentum operator, $E_{F}$ is the Fermi energy, $E$ is the incident electron energy, $v_{F}$ is the Fermi velocity, and $A$ is the magnetic vector potential. MBS (shown in white) occurs in the upper half of the ring at the junction between the superconducting and ferromagnetic layer in the TI (STIM junction) (see Fig. 1) \cite{FuKaneMBSog, fuKaneMBS, nilssonMBS}. The Hamiltonian for the MBS is \cite{colin, nilssonMBS},
\begin{equation} \label{eq2}
H_{M} = -\sigma_{y}E_{M}\, ,
\end{equation}
with $E_{M}$ denoting coupling strength between individual MBS. The STIM junction is connected to the left and right arms of the ring with coupling strengths $\Gamma_{1}$ and $\Gamma_{2}$, respectively. In the next subsection, we outline the scattering via edge modes in the setup and calculate the transmission probability $\mathcal{T}$.

\begin{figure*}
    \centering
    \setlength{\unitlength}{1\textwidth} 
    \begin{picture}(1,0.75) 
        \put(0, 0){\includegraphics[width=1\textwidth]{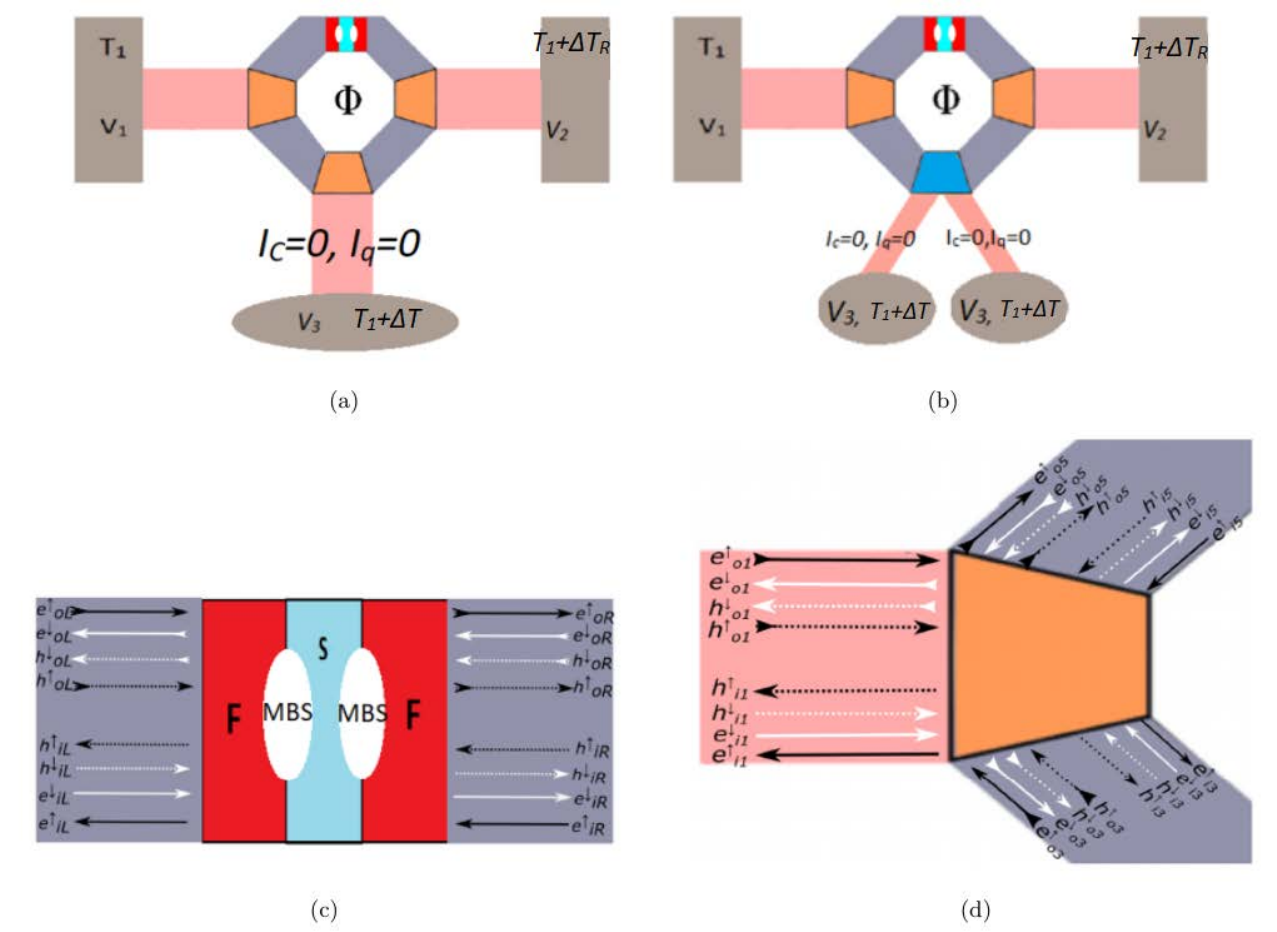}} 
        \put(0.21, 0.65){\makebox(0,0)[c]{\textbf{J1}}} 
        \put(0.325, 0.65){\makebox(0,0)[c]{\textbf{J2}}} 
        \put(0.68, 0.65){\makebox(0,0)[c]{\textbf{J1}}} 
        \put(0.795, 0.65){\makebox(0,0)[c]{\textbf{J2}}} 
        \put(0.82, 0.22){\makebox(0,0)[c]{\huge{\textbf{J1}}}} 
    \end{picture}
    \caption{(a) The 3 terminal Majorana Aharonov-Bohm interferometer with a {BVT} probe, which entails both phase and momentum relaxation \cite{buttikerboth}. A 2D {topological insulator} ring with helical edge modes flowing in the outer and inner edges. The ring is coupled to two regular {topological insulator} leads via couplers {J1 and J2} (shown in orange). The leads connect the AB ring to the reservoirs. A third coupler connects the ring to an inelastic scatterer acting as a {BVT} probe \cite{dvira}, such that the total charge and heat current flowing in the third terminal is zero. The ring is pierced by an Aharonov-Bohm flux. Ferromagnetic (shown in red) and superconducting (shown in cyan) correlations are induced in the top arm of the ring via the proximity effect. (b) The Majorana Aharonov-Bohm interferometer with a {BVT} probe which entails only phase relaxation \cite{PhysRevB.33.3020}, the {BVT} probe consists of two unidirectional probes attached to the bottom via a coupler such that the total charge and heat currents into each probe terminal is zero. The electrons injected into each probe terminal are reinjected back into the setup via the other probe, preserving the momentum. (c) The superconductor-topological insulator-ferromagnet junction. {Majorana bound states} (shown as white ellipses) occurs at the interface of the ferromagnetic (shown in red) and superconducting (shown in cyan) layer. (d) Scattering of the edge modes at the left coupler {J1}. The solid and dashed lines represent the electron and hole edge modes. The black and white lines represent the spin-up and spin-down edge modes. The double-headed arrows represent the outer edge modes, and the single-headed arrows represent the inner edge modes, respectively.}
\end{figure*}

\subsection{Transport in the system via edge modes}
In a 2D quantum Hall ring with an Aharonov-Bohm flux, localized flux-sensitive edge modes develop near the hole, while in the leads (shown in pink in Fig. 1), edge modes are insensitive to flux. To tune the device via an Aharonov-Bohm flux, we need to couple the edge modes in the leads and the edge modes in the ring so that the net conductance is flux-sensitive. It can be achieved via couplers (shown in orange in Fig. 1) in the system that couples the inner and outer edge modes via inter-edge scattering and backscattering. There are three leads in the setup coupled to the ring. The left and right couplers are connected to reservoirs at voltage $V_{1}$ on the left, $V$ on the right, and temperatures $T$ at the left and right reservoirs. A pair of MBS occur in the STIM junction on the top of the ring and act as a backscatter, mixing the electron and hole edge modes. Considering the electron and hole edge modes of spins up and down, we get $8$ edge modes with $4$ edge modes circulating on the outer edge and $4$ edge modes circulating on the inner edge. Since spin-flip scattering does not occur in our setup, we can significantly simplify the calculation by dividing the edge modes into two sets of edge modes of opposite spin that scatter as mirror images of each other. It allows us to calculate the transmission probabilities for a single set and double it to get the net conductance.\\
The first set consists of the spin-up electron and spin-up hole edge modes, and the second set consists of counterpropagating spin-down electron and spin-down hole edge modes. The STIM junction couples the incoming and outgoing edge modes of each set. Among the spin-up {electron and spin-down hole} edge modes, incoming edge modes into the STIM junction are given by $I_{MBS1} = (e^{\uparrow}_{oL}, e^{\uparrow}_{iR}, h^{\downarrow}_{oL}, h^{\downarrow}_{iR})$ and the outgoing edge modes are given by $O_{MBS1} = (e^{\uparrow}_{oR}, e^{\uparrow}_{iL}, h^{\downarrow}_{iL}, h^{\downarrow}_{oR})$ (see Fig. 1 (b)). Similarly, among the spin-down {electron and spin-up hole} edge modes, the incoming edge modes are $I_{MBS2} = (e^{\downarrow}_{oR}, e^{\downarrow}_{iL}, h^{\uparrow}_{iL}, h^{\uparrow}_{oR})$, while the outgoing edge modes are $O_{MBS2} = (e^{\downarrow}_{oL}, e^{\downarrow}_{iR}, h^{\uparrow}_{oL}, h^{\uparrow}_{iR})$. {$o$ and $i$ stand for the outer and inner edge modes, respectively, whereas $L$ and $R$ stand for the left and right terminals. }The scattering in each case is the exact mirror image of the other. We can relate the incoming and outgoing edge modes using a $4 \times 4$ scattering matrix $S_{MBS}$ such that $O_{MBSi} = S_{MBS}I_{MBSi}, i \in \{1,2\}$, where $S_{MBS}$ is given by \cite{nilssonMBS, colin}

\begin{subequations} \label{eq3}
\begin{equation} \label{eq3a}
S_{MBS} = \begin{pmatrix}
1 + ix & -y & ix & -y\\
y & 1 + ix' & y & ix'\\
ix & -y & 1 + ix & -y\\
y & ix' & y & 1 + ix'
\end{pmatrix}
\end{equation}
where,
\begin{equation} \label{eq3b}
\begin{split}
x =\frac{\Gamma_{1}(E + i\Gamma_{2})}{z},x' = \frac{\Gamma_{2}(E + i\Gamma_{1})}{z},\\
y = \frac{E_{M}\sqrt{\Gamma_{1}\Gamma_{2}}}{z},
z=E_{M}^{2}-(E + i\Gamma_{1})(E + i\Gamma_{2})\, 
\end{split}
\end{equation}
\end{subequations}
and $\Gamma_{1}$, $\Gamma_{2}$ are the strengths of the couplers coupling the MBS to the left and right arms of the upper ring.\\
The couplers (see Fig. 1 (d)) couple the inner and outer edge modes via backscattering and the ring to the leads. In the left coupler, the incoming  edge modes {for spin-up electron and spin-down hole} are given by $I_{1} = (e^{\uparrow}_{o1},h^{\downarrow}_{o1},e^{\uparrow}_{o3},h^{\downarrow}_{o3},e^{\uparrow}_{i5},h^{\downarrow}_{i5})$, and the corresponding outgoing edge modes are given by $O_{1} = (e^{\uparrow}_{i1},h^{\downarrow}_{i1},e^{\uparrow}_{i3},h^{\downarrow}_{i3},e^{\uparrow}_{o5},h^{\downarrow}_{o5})$. Similarly, the incoming edge modes {for spin-down electron and spin-up} hole are given by $I_{2}=(e^{\downarrow}_{i1},h^{\uparrow}_{i1},e^{\downarrow}_{i3},h^{\uparrow}_{i3},e^{\downarrow}_{o5},h^{\uparrow}_{o5})$, and the corresponding outgoing spin-down edge modes are $O_{2} = (e^{\downarrow}_{o1},h^{\uparrow}_{o1},e^{\downarrow}_{o3},h^{\uparrow}_{o3},e^{\downarrow}_{i5},h^{\uparrow}_{i5})$. The scattering matrix for the couplers is a $6 \times 6$ matrix $S$ such that $O_{i} = SI_{i}, i \in \{1,2\}$ is given by \cite{buttikerboth}
\begin{equation} \label{eq4}
S =
\begin{pmatrix}
-\left(\sqrt{1 - 2 \alpha}\right)I && \sqrt{\alpha}I && \sqrt{\alpha}I\\
& & & &\\
\sqrt{\alpha}I && \frac{1}{2}\left(\sqrt{1 - 2\alpha} - 1\right)I && \frac{1}{2}\left(\sqrt{1 - 2\alpha} - 1\right)I\\
& & & \\
\sqrt{\alpha}I && \frac{1}{2}\left(\sqrt{1 - 2\alpha} + 1\right)I && \frac{1}{2}\left(\sqrt{1 - 2\alpha} - 1\right)I
\end{pmatrix},
\end{equation}

where $I$ is the $2 \times 2$ identity matrix. {In our calculation, we take $\alpha = \frac{1}{2}$, however to analyze the plots of elastic scattering at couplers in Analysis in Sec. VI, we also consider $\alpha = \frac{4}{9}$ too in Eq. 4.} While traversing the ring, the spin-up electrons and holes in the edge modes acquire a propagating phase \cite{colin} as follows:\\
in the upper arm, left of the STIM junction:

\begin{equation} \label{eq5}
\begin{split}
e^{\uparrow}_{i5} = e^{ik_{e}l_{1}}e^{\frac{-i\phi l_{1}}{L}} e^{\uparrow}_{iL},
e^{\uparrow}_{oL} = e^{ik_{e}l_{1}}e^{\frac{i\phi l_{1}}{L}} e^{\uparrow}_{o5}, \\
h^{\downarrow}_{i5} = e^{ik_{h}l_{1}}e^{\frac{i\phi l_{1}}{L}} h^{\downarrow}_{iL},
h^{\downarrow}_{oL} = e^{ik_{h}l_{1}}e^{\frac{-i\phi l_{1}}{L}} h^{\downarrow}_{o5},
\end{split}
\end{equation}
for the upper arm, right of STIM junction:
\begin{equation} \label{eq6}
\begin{split}
e^{\uparrow}_{o6} = e^{ik_{e}l_{2}}e^{\frac{i\phi l_{2}}{L}} e^{\uparrow}_{oR},
e^{\uparrow}_{iR} = e^{ik_{e}l_{2}}e^{\frac{-i\phi l_{2}}{L}} e^{\uparrow}_{i6},\\
h^{\downarrow}_{o6} = e^{ik_{h}l_{2}}e^{\frac{-i\phi l_{2}}{L}} h^{\downarrow}_{oR},
h^{\downarrow}_{iR} = e^{ik_{h}l_{2}}e^{\frac{i\phi l_{2}}{L}} h^{\downarrow}_{i6},
\end{split}
\end{equation} 
such that $l_{1} + l_{2} = l_{u}$, i.e. length of the upper arm of the ring. For the lower arm of the ring, left of the {BVT} probe:
\begin{equation} \label{eq7}
\begin{split}
e^{\uparrow}_{o3} = e^{ik_{e}l'_{1}}e^{\frac{i\phi l'_{2}}{L}} e^{\uparrow}_{o4},
e^{\uparrow}_{i4} = e^{ik_{d}l'_{1}}e^{\frac{-i\phi l'_{2}}{L}} e^{\uparrow}_{i3},\\
h^{\downarrow}_{o3} = e^{ik_{h}l'_{1}}e^{\frac{-i\phi l'_{2}}{L}} h^{\downarrow}_{o4},
h^{\downarrow}_{i4} = e^{ik_{h}l'_{1}}e^{\frac{i\phi l'_{2}}{L}} h^{\downarrow}_{i3},
\end{split}
\end{equation}
for the lower arm of the ring, right of the {BVT} probe:
\begin{equation} \label{eq8}
\begin{split}
e^{\uparrow}_{o4} = e^{ik_{e}l'_{1}}e^{\frac{i\phi l'_{2}}{L}} e^{\uparrow}_{o5},
e^{\uparrow}_{i5} = e^{ik_{d}l'_{1}}e^{\frac{-i\phi l'_{2}}{L}} e^{\uparrow}_{i4},\\
h^{\downarrow}_{o4} = e^{ik_{h}l'_{1}}e^{\frac{-i\phi l'_{2}}{L}} h^{\downarrow}_{o5},
h^{\downarrow}_{i5} = e^{ik_{h}l'_{1}}e^{\frac{i\phi l'_{2}}{L}} h^{\downarrow}_{i4},
\end{split}
\end{equation}
such that $l'_{1} + l'_{2} = l_{d}$, i.e., the length of the bottom arm of the ring. The total length of the ring $L$ is given by $L = l_{u} + l_{d}$. $k_{e} = (E + E_{f})/\hbar v_{F}$, and $k_{h} = (E - E_{f})/\hbar v_{F}$ are electron and hole wave vectors in the 2D TI. $\phi$ is the Aharonov-Bohm flux taken in units of the flux quantum $\phi_{0} = hc/e$. One can similarly find the phase acquired by the spin-down electrons and holes, which is the exact mirror image of spin-up electrons and holes. In the next section, we describe thermoelectric transport in multi-terminal mesoscopic systems and introduce inelastic scattering by adding a {BVT} probe \cite{buttikerboth, PhysRevB.33.3020, dvira} to the Majorana ABI.
\section{Inelastic scattering and Buttiker voltage temperature probe}
We explain the thermoelectric transport in our setup shown in Fig. 1 using Onsager relations and Landauer-Buttiker scattering theory \cite{Buttiker1983365, PhysRevB.53.16390, lambert1993multi, beneticasati}. We denote the charge and heat currents in the $i^{th}$ terminal by current vector $I^{s}_{i} = (I^{s}_{ic}, I^{s}_{iq}),$ where $I^{s}_{ic} = I^{s,e}_{ic} + I^{s,h}_{ic}$ is the total charge current, and $I^{s}_{iq} = I^{s,e}_{iq} + I^{s,h}_{iq}$ the total heat current with spin $s \in \{ \uparrow, \downarrow \}$ electrons and holes. The current vector {$I^{s,k}_{i} = (I^{s,k}_{ic}, I^{s,k}_{iq})^{T}$, with $k \in \{e, h\}$} can be related to force vector $F_{ij} = (V_{j}, \Delta T_{ij})^{T}$ (where $V_{j}$ is the voltage at the $j^{th}$ terminal, and $\Delta T_{ij} = T_{j} - T_{i}$ is the temperature difference across the $i^{th}$ and the $j^{th}$ terminals) by the Onsager matrix $L^{sr;kl}_{ij}$ such that {$I^{s,k}_{i} = \sum_{j} L_{ij}^{s,k} F_{ij}$}, where \cite{PhysRevB.53.16390, lambert1993multi},
\begin{equation} \label{eq9}
\begin{split}
{L^{s;k}_{ij} = \begin{pmatrix}
L^{s;k}_{ij;cV} & L^{s;k}_{ij;cT}\\
L^{s;k}_{ij;qV} & L^{s;k}_{ij;qT}
\end{pmatrix}}
{= \frac{1}{h}\int^{\infty}_{-\infty} dE (\delta_{ij} - \mathcal{T}^{ss;kk}_{ij}(E, E_{F})}\\ {+ \mathcal{T}^{rs;lk}_{ij}(E, E_{F})) 
\times \xi(E, E_{F})L_{0}(E, E_{F}), \hspace{0.2cm} \text{with} \hspace{0.1cm} r \ne s \hspace{0.1cm} \text{and} \hspace{0.1cm} l \ne k}
\end{split}
\end{equation}

\begin{equation} \label{eq10}
\text{ with $L_{0}(E, E_{F}) = G_{0}$}
\begin{pmatrix}
1 & (E - E_{F})/eT_{i}\\
(E-E_{F})/e & (E-E_{F})^{2}/e^{2}T_{i}
\end{pmatrix},
\end{equation}

where $\delta_{ij}$ is the Dirac Delta function, $f(E, E_{F})$ being the Fermi function $(f(E, E_{F}) = \frac{1}{1 + e^{(E-E_{F})/k_{B}T_{i}}}$), with $\xi(E, E_{F}) =\frac{-\partial f(E, E_{F})}{\partial E}$, $T_1$ and {$T_1 + \Delta T_R$} are the temperatures of the left and right reservoirs. In contrast, {$T_1 + \Delta T$} is the temperature of the {BVT} probe, $k_{B}$ is the Boltzmann constant, $G_{0} = (e^{2}/\hbar)$ is the conductance quantum, $E$ is particle energy, $E_{F}$ is Fermi energy, and $h$ is Planck's constant. The Onsager matrix elements describe the thermal and electrical response to the voltage and temperature bias. {$L^{s;k}_{ij;cV}$}, is the electrical conductance, {$L^{s;k}_{ij;cT}$} is the electrical response to the temperature difference, {$L^{s;k}_{ij;qV}$} is the thermal response to the voltage difference, and {$L^{s;k}_{ij;qT}$} is the thermal response generated due to the temperature difference due to a spin $s$ electron or hole ($s \in \{ \uparrow, \downarrow \}$) being transmitted from the $j^{th}$ terminal into the $i^{th}$ terminal as a spin $s$ electron or hole ($k \in \{ e, h \}$). {$\mathcal{T}_{ij}^{ss,kk}$ is the transmission probability for a particle of type $k \in \{e, h \}$ with spin $s \in \{\uparrow, \downarrow\}$ to transmit from terminal $j$ to terminal $i$ as the same particle type $k$ and same spin $s$. Similarly, $\mathcal{T}_{ij}^{sr,lk}$ is the transmission probability for a particle of type $k \in \{e, h \}$ with spin $s \in \{\uparrow, \downarrow\}$ to transmit from terminal $j$ to terminal $i$ as the particle type $l \ne k$ and spin $r \ne s$.} MBS is a superposition of electrons and holes of the same spin; thus, the scattering due to MBS can cause electron-electron scattering with {same} spin or electron-hole scattering between particles of the {opposite} spin. Since the opposite spin edge modes are mirror images of each other and there is no spin-flip scattering in our system (see Fig. 1), we can write {$\mathcal{T}^{\uparrow \uparrow; kk}_{ij} = \mathcal{T}^{\downarrow \downarrow; kk}_{ij}$, and $\mathcal{T}^{\downarrow \uparrow; lk}_{ij} = \mathcal{T}^{\uparrow \downarrow; kl}_{ij}$ and $\mathcal{T}^{\uparrow \downarrow}_{ij;kk} = \mathcal{T}^{\uparrow \downarrow}_{ij;kk} = 0$.} From Eq. (\ref{eq9}), we can write:
\begin{subequations} \label{eq11}
\begin{equation} \label{eq11a}
\begin{split}
{I^{\uparrow, e}_{ic} = \sum_{j} L^{\uparrow; e}_{ij;cV}V_{j} + \sum_{j} L^{\uparrow; e}_{ij;cT}\Delta T_{ij},} \\
\text{ and } {I^{\uparrow, e}_{iq} = \sum_{j} L^{\uparrow; e}_{ij;qV}V_{j} + \sum_{j} L^{\uparrow; e}_{ij;qT}\Delta T_{ij},} \\
\end{split}
\end{equation}
\begin{equation} \label{eq11b}
\begin{split}
{I^{\downarrow, e}_{ic} = \sum_{j} L^{\downarrow; e}_{ij;cV}V_{j} + \sum_{j} L^{\downarrow; e}_{ij;cT}\Delta T_{ij},} \\
\text{ and } {I^{\downarrow, e}_{iq} = \sum_{j} L^{\downarrow; e}_{ij;qV}V_{j} + \sum_{j} L^{\downarrow; e}_{ij;qT}\Delta T_{ij},} \\
\end{split}
\end{equation}
\begin{equation} \label{eq11c}
\begin{split}
{I^{\uparrow, h}_{ic} = \sum_{j} L^{\uparrow; h}_{ij;cV}V_{j} + \sum_{j} L^{\uparrow; h}_{ij;cT}\Delta T_{ij},} \\
\text{ and } {I^{\uparrow, h}_{iq} = \sum_{j} L^{\uparrow; h}_{ij;qV}V_{j} + \sum_{j} L^{\uparrow; h}_{ij;qT}\Delta T_{ij},} \\
\end{split}
\end{equation}
\begin{equation} \label{eq11d}
\begin{split}
{I^{\downarrow, h}_{ic} = \sum_{j} L^{\downarrow; h}_{ij;cV}V_{j} + \sum_{j} L^{\downarrow; h}_{ij;cT}\Delta T_{ij},} \\
\text{ and } {I^{\downarrow, h}_{iq} = \sum_{j} L^{\downarrow; h}_{ij;qV}V_{j} + \sum_{j} L^{\downarrow; h}_{ij;qT}\Delta T_{ij}.} \\
\end{split}
\end{equation}
\end{subequations}
This section will introduce inelastic scattering in our system using a {BVT} probe \cite{dvira}. In the setup shown in Fig. 1, an additional lead is connected to the ABI via a coupler that connects the ABI to a {BVT} probe such that the total charge current and the total heat current flowing into terminal 3 are zero. The S-matrices for the left and right couplers are described in Eq. (\ref{eq4}), and they scatter electrons/holes elastically. The {BVT} probe induces inelastic scattering by setting the total charge and heat current passing through itself to zero. It ensures that no net current flows out of the setup into the probe (in section IV, analysis, we will look at the scaling of the Lorenz ratio with respect to the strength of inelastic scattering in the single-particle picture). We study two types of inelastic scattering in our setup, with only phase relaxation \cite{PhysRevB.33.3020} and with both phase relaxation and momentum relaxation \cite{buttikerboth}. We compare and contrast our result with that obtained using Luttinger formalism \cite{buccheri}.

\subsection{Inelastic scattering with both phase and momentum relaxation}
In the first model (see Fig. 1 (a)), we use an inelastic scatterer with both phase and momentum relaxation \cite{buttikerboth}. The S-matrix for the inelastic scatterer with both phase and momentum relaxation is given below,
\begin{equation} \label{eq12}
S =
\begin{pmatrix}
-(p+q)I & \sqrt{\epsilon}I & \sqrt{\epsilon}I\\
\sqrt{\epsilon}I & pI & qI\\
\sqrt{\epsilon}I & qI & pI
\end{pmatrix},
\end{equation}
{where $p = \frac{1}{2} \left(\sqrt{1 - 2 \epsilon} - 1\right)$ and $p = \frac{1}{2} \left(\sqrt{1 - 2 \epsilon} + 1\right)$, see also Ref. \cite{buttikerboth}.} {When $\epsilon = 0$, then complete backscattering occurs and the Aharanov-Bohm ring is disconnected from the Buttiker probe. It means no electron has entered into the Buttiker probe from Aharanov-Bohm ring, which indicates zero inelastic scattering.} {The {BVT} probe is a phenomenological method used to incorporate the effects of inelastic scattering in mesoscopic samples. Its primary role is to provide a theoretical foundation, using the single-electron transport-based Landauer-Buttiker model, to explain experimental results where electron-electron and electron-phonon scattering occur. This approach addresses the question of whether one can gain insight into how inelastic effects influence electron transport without performing an exact many-body calculation. This was the motivation behind Buttiker's introduction of the voltage probe model, as discussed in Refs. \cite{buttikerboth, PhysRevB.33.3020}. In this model, it is ensured that the charge current vanishes at the voltage probe. Electrons lose their phase memory upon entering the voltage probe and are re-injected back into the sample with the same probability. Later, M. Kilgour and D. Segal extended this concept by introducing the {BVT} probe, as detailed in Refs. \cite{dvira, kilgour2015charge, kilgour2015tunneling}. This generalization of the Buttiker voltage probe to thermoelectric transport ensures that both charge and heat currents vanish at the {BVT} probe. Experimentally, a voltage probe can be implemented by ensuring a purely coherent system. Additionally, the system must be tuned to the linear response regime (i.e., near equilibrium), which involves applying small bias voltages (on the order of millivolts) and small temperature biases (around 1 Kelvin). Experimentally, one can also vary the inelastic scattering parameter $\epsilon$ in a BVT as has been performed in \cite{PhysRevLett.102.236802}.}

 We can calculate the charge and heat currents for the S-matrix given in Eq. (\ref{eq12}) by using the respective S-matrix for the {BVT} probe \cite{ dvira} after solving for the transmission probabilities {$\mathcal{T}^{ss;kk}_{ij}$ and $\mathcal{T}^{rs;lk}_{ij}$}, {see, Appendix A}. The total charge current in the {BVT} probe is thus given by,
\begin{equation} \label{eq25}
I_{3c} = I^{\uparrow e}_{3c} + I^{\uparrow h}_{3c} + I^{\downarrow e}_{3c} + I^{\downarrow h}_{3c},
\end{equation}
and the total heat current is given by,
\begin{equation} \label{eq26}
I_{3q} = I^{\uparrow e}_{3q} + I^{\uparrow h}_{3q} + I^{\downarrow e}_{3q} + I^{\downarrow h}_{3q}.
\end{equation}
In order to introduce inelastic scattering via the {BVT} probe, we should have $I_{3c}(V, T + \Delta T) = I_{3q}(V, T + \Delta T) = 0$. In our setup, we measure the charge and heat currents in the second terminal. The total charge current in the second terminal is given by,
\begin{equation} \label{eq27}
I_{2c} = I^{\uparrow e}_{2c} + I^{\uparrow h}_{2c} + I^{\downarrow e}_{2c} + I^{\downarrow h}_{2c},
\end{equation}
and the total heat current is given by,
\begin{equation} \label{eq28}
I_{2q} = I^{\uparrow e}_{2q} + I^{\uparrow h}_{2q} + I^{\downarrow e}_{2q} + I^{\downarrow h}_{2q}.
\end{equation}
In order to find the thermoelectric coefficients and, subsequently, the Lorenz ratio, we write $I_{2q}$ and $I_{2c}$ in terms of $V$ and $\Delta T$  by eliminating $V_{3}$ and $\Delta T$ from Eqs. (\ref{eq18}-\ref{eq21}). We use the condition of the {BVT} probe ($I_{3c}(V, T + \Delta T) = I_{3q}(V, T + \Delta T) = 0$) in order to write all the voltages and temperatures in terms of $V$ and $\Delta T$. Thus, $I_{2c}$ and $I_{2q}$ can now be written solely in terms of $V$ and $\Delta T$ as,
\begin{equation} \label{eq29}
I_{2c} = L'_{cV}V + L'_{cT}\Delta T_R
\text{ and } I_{2q} = L'_{qV}V + L'_{qT}\Delta T_R,
\end{equation}
where $L'_{cV}, L'_{qV}, L'_{cT}, L'_{qT}$ are the effective Onsager coefficients for conduction in the second terminal and are functions of {$G(\mathcal{T}^{s;k}_{ij}(E))$, $S(\mathcal{T}^{s;k}_{ij}(E))$, $L(\mathcal{T}^{s;k}_{ij}(E))$} as defined in Eqs. (\ref{eq22}-\ref{eq24}). $L'_{cV}$ is the effective electrical conductance in the second terminal given as,
$
L'_{cV} = \frac{I_{2c}}{V}|_{\Delta T = 0}.
$
$L'_{cT}$ is the electrical response to the temperature difference and is given by $L'_{cT} = \frac{I_{2c}}{\Delta T_R}|_{V = 0}$. Similarly, the thermal response to the voltage difference $L'_{qV}$ is given by $L'_{qV} = \frac{I_{2q}}{V}|_{\Delta T_R = 0}$, and the thermal response to the temperature difference is given by $L'_{qT} = \frac{I_{2q}}{\Delta T}|_{V = 0}$. The effective Onsager coefficients are then given as,

\begin{widetext}
\begin{equation} \label{eq30}
{L'_{cV} = \sum_{s,k} \left [G_{22}^{s,k} + \delta(G_{23}^{s,k}(S_{33}^{s,k} S_{32}^{s,k} - G_{32}^{s,k}L_{33}^{s,k}) - S_{23}^{s,k}(G_{33}^{s,k}S_{32}^{s,k}-G_{32}^{s,k}S_{33}^{s,k}))\right]},
\end{equation}
\begin{equation} \label{eq31}
{L'_{cT} = \sum_{s,k} \left [S_{22}^{s,k} + \delta(G_{23}^{s,k}(S_{33}^{s,k} L_{32}^{s,k} - S_{32}^{s,k}L_{33}^{s,k}) - S_{23}^{s,k}(G_{33}^{s,k}L_{32}^{s,k}-L_{32}^{s,k}S_{33}^{s,k}))\right]},
\end{equation}

\begin{equation} \label{eq32}
{L'_{qV} = \sum_{s,k} \left [S_{22}^{s,k} + \delta(S_{23}^{s,k}(S_{33}^{s,k} S_{32}^{s,k} - G_{32}^{s,k}L_{33}^{s,k}) - L_{23}^{s,k}(G_{33}^{s,k}S_{32}^{s,k}-G_{32}^{s,k}S_{33}^{s,k}))\right]},
\end{equation}

\begin{equation} \label{eq33}
{L'_{qT} = \sum_{s,k} \left [L_{22}^{s,k} + \delta(S_{23}^{s,k}(S_{33}^{s,k} L_{32}^{s,k} - S_{32}^{s,k}L_{33}^{s,k}) - L_{23}^{s,k}(G_{33}^{s,k}L_{32}^{s,k}-S_{32}^{s,k}S_{33}^{s,k}))\right]},
\end{equation}
\end{widetext}
where {$\delta = \dfrac{1}{\sum_{s,k} \left[G_{33}^{s,k}L_{33}^{s,k} - (S_{33}^{s,k})^2 \right]} $}. The effective electrical conductance is the total charge current generated due to the voltage difference and is given by $\sigma' = L'_{cV}$. The effective thermal conductance \cite{strainedrefg} is the heat current generated by a unit temperature bias without any charge current. The effective thermal conductance of the setup with {Eq. (\ref{eq12})} is then given by,
\begin{equation} \label{eq34}
\kappa' = \frac{I_{2q}}{\Delta T}|_{I_{ic} = 0} =\frac{L'_{cV}L'_{qT} - L'_{cT}L'_{qV}}{L'_{cV}}.
\end{equation}

Wiedemann-Franz (WF) law states that the ratio of the effective thermal conductance ($\kappa'$) to the effective electric conductance ($\sigma'$) is proportional to the temperature \cite{weideman} of the TI ($T$). We define the Lorenz ratio $W$ as,
\begin{equation} \label{eq35}
W = \frac{\kappa'}{\sigma'}.
\end{equation}
When WF law is preserved, $W = W_{0}$, where $W_{0} = \frac{\kappa_{0}}{G_{0}}$, where $G_{0} = \dfrac{e^{2}}{\hbar}$ and $\kappa_{0} = \pi^{2} k_{B}^{2} T/3h$, where $T$ is temperature of the reservoirs 1 and 2. WF law is derived from the fact that in condensed matter systems, both charge and heat are carried by the quasiparticles in the conductor, namely, the electrons and holes. Quasiparticles in conductors generally follow particle-hole symmetry (PHS), i.e., the electron energy levels are symmetric to the hole energy levels. When PHS is preserved in the system, WF law is preserved, and the Lorenz ratio is not violated \cite{weideman,ramosweideman}. The breakdown of PHS causes violations of WF law. In the next subsection, we include inelastic scattering with the help of the {BVT} probe but with phase relaxation only.

\subsection{Inelastic scattering with only phase relaxation}
Fig. 1 (b) shows the MBS ABI with only phase relaxation \cite{PhysRevB.33.3020}. The third terminal, i.e., the {BVT} probe, is divided into two terminals that are connected to reservoirs at voltage $V_{3}$ and temperature $T + \Delta T$ such that the total charge current and heat current flowing in each terminal is zero. The inelastic scatterers are connected via a coupler described by a $ 4 \times 4$ scattering matrix given by \cite{PhysRevB.33.3020},
\begin{equation} \label{eq36}
S_{p} =
\begin{pmatrix}
0 & \sqrt{1 - \epsilon}I & 0 & -\sqrt{\epsilon}I\\
\sqrt{1 - \epsilon}I & 0 & -\sqrt{\epsilon}I & 0\\
\sqrt{\epsilon}I & 0 & \sqrt{1 - \epsilon}I & 0\\
0 & \sqrt{\epsilon}I & 0 & \sqrt{1 - \epsilon}I
\end{pmatrix},
\end{equation}
where $\epsilon \in [0, 1]$, and $I$ is the $2 \times 2$ scattering matrix. $\epsilon = 1$ denotes maximal coupling, while $\epsilon \to 0$ denotes no coupling \cite{PhysRevB.33.3020}.  Similar to the derivation of the Onsager matrix elements for the inelastic scatterer with both phase and momentum relaxation, one can find the Onsager matrix elements for the inelastic scatterer with only phase relaxation, {see Appendix B}. First, we write {charge and heat current} in Eqs. (\ref{eq41}-\ref{eq44}) in terms of $V$, and $\Delta T_R$ only by using the condition of the {BVT} probe ($I_{3c}(V, T + \Delta T) = I_{3q}(V, T + \Delta T) = I_{4c}(V, T + \Delta T) = I_{4q}(V, T + \Delta T) = 0$). This allows us to write $I_{2c}$, and $I_{2q}$ in terms of $V$ and $\Delta T$ only as,
\begin{equation} \label{eq45}
\begin{split}
I_{2c} = L''_{cV}V + L''_{cT}\Delta T_R
\text{, and } I_{2q} = L''_{qV}V + L''_{qT}\Delta T_R. \\
\end{split}
\end{equation}
In Eq.~(\ref{eq45}) $L''_{cV}, L''_{cT}, L''_{qV}, L''_{qT}$ are the effective Onsager matrix elements for the setup with phase relaxation only (see Fig. 1 (b)) and are given by,

\begin{widetext}

\begin{equation} \label{eq46}
\begin{split}
{L''_{cV} = \sum_{s,k} \bigg[G_{22}^{s,k} + \frac{1}{\Delta} (G_{23}^{s,k} + G_{24}^{s,k})((S_{33}^{s,k} + S_{34}^{s,k} + S_{43}^{s,k} + S_{44}^{s,k})(S_{32}^{s,k}+S_{42}^{s,k}) - (G_{32}^{s,k} + G_{42}^{s,k})(L_{33}^{s,k} + L_{34}^{s,k} + L_{43}^{s,k} + L_{44}^{s,k}))}\\ {- \frac{1}{\Delta}(S_{23}^{s,k} + S_{24}^{s,k})((G_{33}^{s,k} + G_{34}^{s,k} + G_{43}^{s,k} + G_{44}^{s,k})(S_{32}^{s,k}+S_{42}^{s,k})-(G_{32}^{s,k} + G_{42}^{s,k})(S_{33}^{s,k} + S_{34}^{s,k} + S_{43}^{s,k} + S_{44}^{s,k}) \bigg]}
\end{split}
\end{equation}

\begin{equation} \label{eq47}
\begin{split}
{L''_{cT} = \sum_{s,k} \bigg[S_{22}^{s,k} + \frac{1}{\Delta} (G_{23}^{s,k} + G_{24}^{s,k})((S_{33}^{s,k} + S_{34}^{s,k} + S_{43}^{s,k} + S_{44}^{s,k})(L_{32}^{s,k}+L_{42}^{s,k}) - (S_{32}^{s,k} + S_{42}^{s,k})(L_{33}^{s,k} + L_{34}^{s,k} + L_{43}^{s,k} + L_{44}^{s,k}))}\\ {- \frac{1}{\Delta}(S_{23}^{s,k} + S_{24}^{s,k})((G_{33}^{s,k} + G_{34}^{s,k} + G_{43}^{s,k} + G_{44}^{s,k})(L_{32}^{s,k}+L_{42}^{s,k})-(S_{32}^{s,k} + S_{42}^{s,k})(S_{33}^{s,k} + S_{34}^{s,k} + S_{43}^{s,k} + S_{44}^{s,k}) \bigg]}
\end{split}
\end{equation}

\begin{figure*}
     \centering
     \begin{subfigure}[b]{0.40\textwidth}
         \centering
         \includegraphics[width=\textwidth]{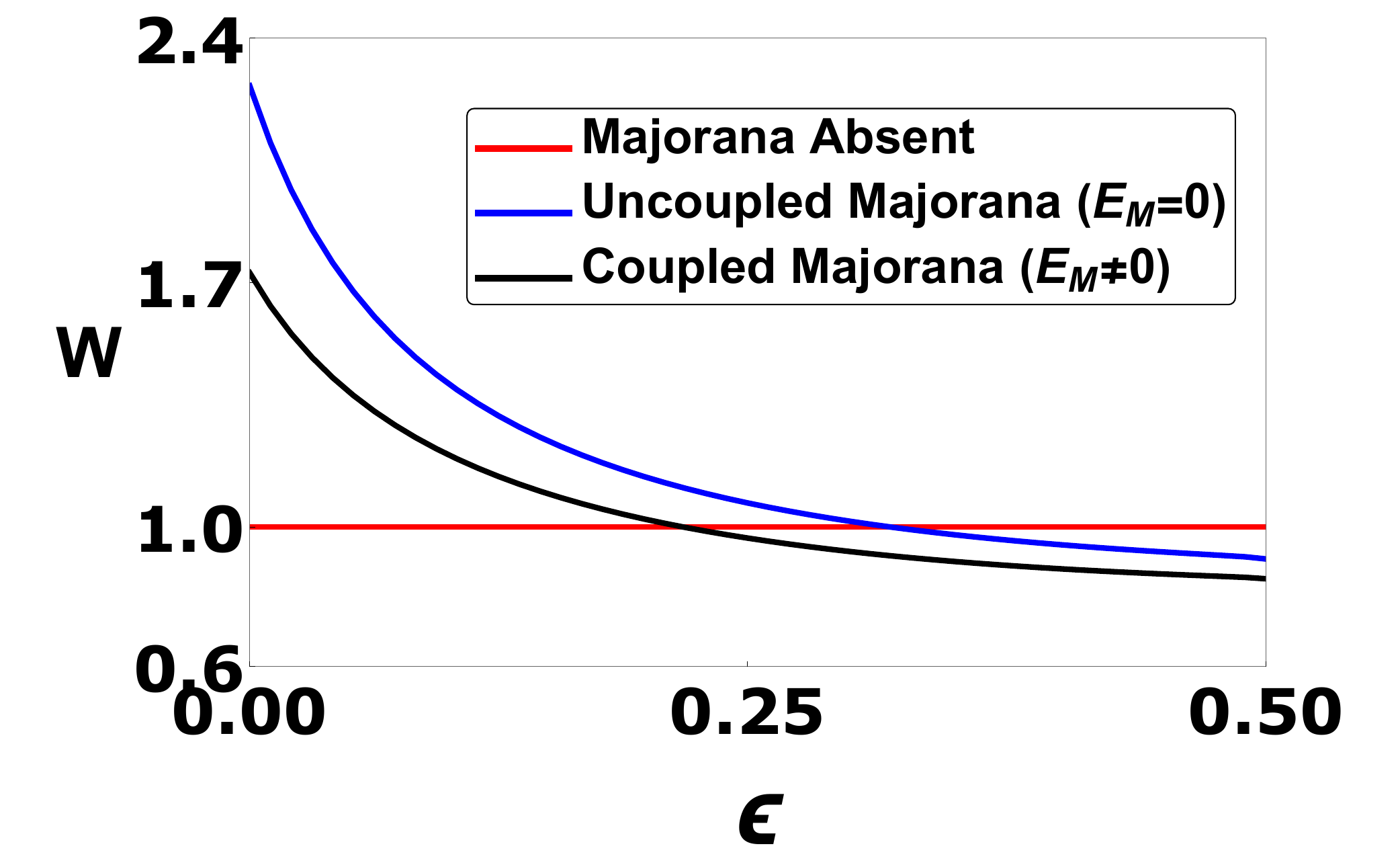}
         \caption{}
     \end{subfigure}
     \hspace{0.05cm}
     \begin{subfigure}[b]{0.40\textwidth}
         \centering
         \includegraphics[width=\textwidth]{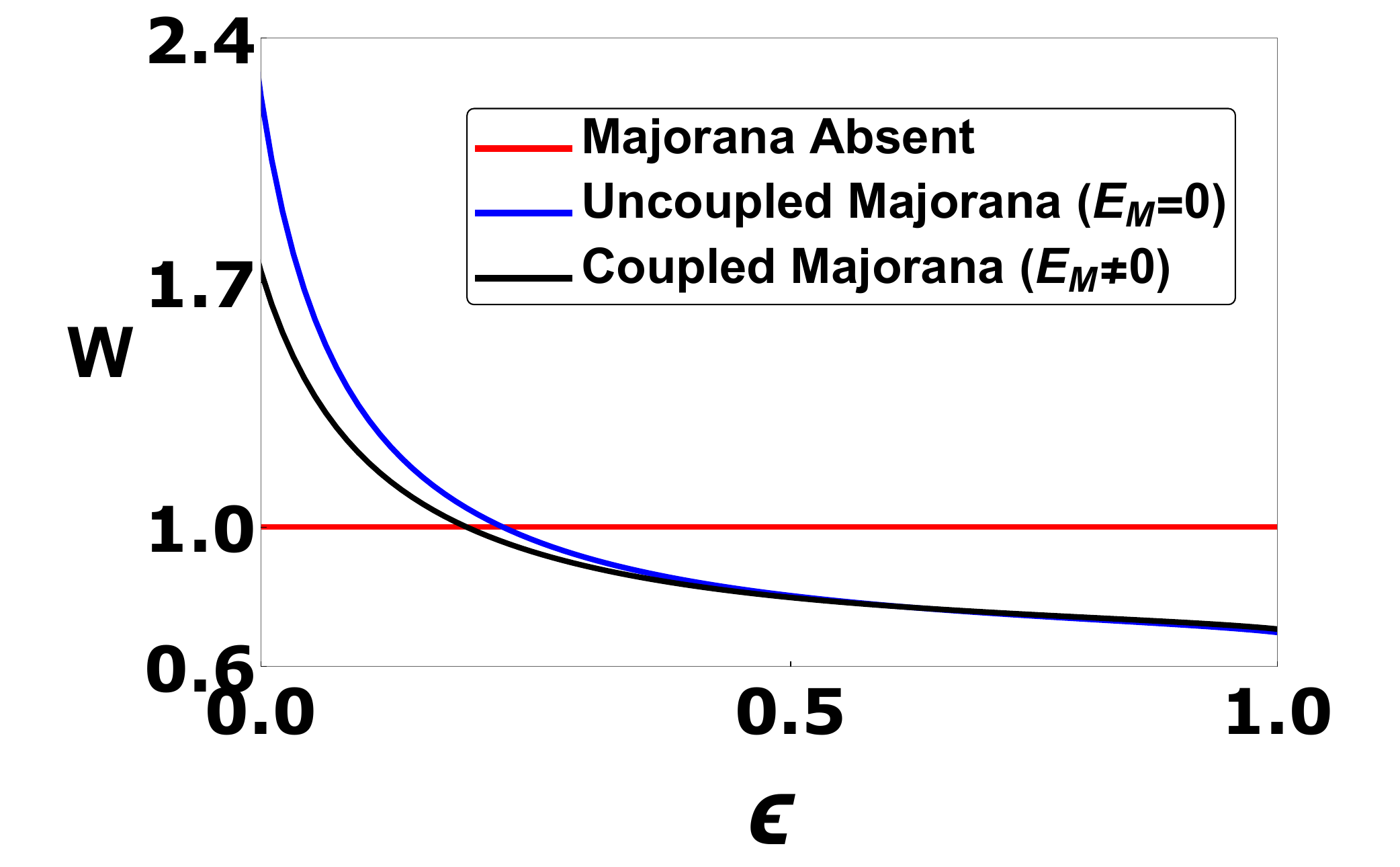}
         \caption{}
     \end{subfigure}
        \caption{{Lorentz ratio ($W$) vs. the coupling strength ($\epsilon$) of the {BVT} probe for (a) with both phase and momentum relaxation (b) with phase relaxation. Parameters are $E_F = 10 \mu eV$, $E_M = 0.3 \mu eV$, $\Gamma = 1 \mu eV$, $\phi = \phi_0$ and elastic scattering at junction $J_1, J_2$, where $\phi = h c / e$ and $\alpha = \frac{1}{2}$ and represents elastic scattering at J1 and J2.}}
         \label{fig:12}
       \end{figure*}

\begin{equation} \label{eq48}
\begin{split}
{L''_{qV} = \sum_{s,k} \bigg[S_{22}^{s,k} + \frac{1}{\Delta} (S_{23}^{s,k} + S_{24}^{s,k})((S_{33}^{s,k} + S_{34}^{s,k} + S_{43}^{s,k} + S_{44}^{s,k})(S_{32}^{s,k}+ S_{42}^{s,k}) - (G_{32}^{s,k} + G_{42}^{s,k})(L_{33}^{s,k} + L_{34}^{s,k} + L_{43}^{s,k} + L_{44}^{s,k}))}\\ {- \frac{1}{\Delta}(L_{23}^{s,k} + L_{24}^{s,k})((G_{33}^{s,k} + G_{34}^{s,k} + G_{43}^{s,k} + G_{44}^{s,k})(S_{32}^{s,k}+S_{42}^{s,k})-(G_{32}^{s,k} + G_{42}^{s,k})(S_{33}^{s,k} + S_{34}^{s,k} + S_{43}^{s,k} + S_{44}^{s,k}) \bigg]}
\end{split}
\end{equation}

\begin{equation} \label{eq49}
\begin{split}
{L''_{qT} = \sum_{s,k} \bigg[L_{22}^{s,k} + \frac{1}{\Delta} (S_{23}^{s,k} + S_{24}^{s,k})((S_{33}^{s,k} + S_{34}^{s,k} + S_{43}^{s,k} + S_{44}^{s,k})(L_{32}^{s,k}+L_{42}^{s,k}) - (S_{32}^{s,k} + S_{42}^{s,k})(L_{33}^{s,k} + L_{34}^{s,k} + L_{43}^{s,k} + L_{44}^{s,k}))}\\ {- \frac{1}{\Delta}(L_{23}^{s,k} + L_{24}^{s,k})((G_{33}^{s,k} + G_{34}^{s,k} + G_{43}^{s,k} + G_{44}^{s,k})(L_{32}^{s,k}+L_{42}^{s,k})-(S_{32}^{s,k} + S_{42}^{s,k})(S_{33}^{s,k} + S_{34}^{s,k} + S_{43}^{s,k} + S_{44}^{s,k}) \bigg]}
\end{split}
\end{equation}
where, $\Delta = \sum_{s,k} \bigg[(G_{33}^{s,k} + G_{34}^{s,k} + G_{43}^{s,k} + G_{44}^{s,k}) (L_{33}^{s,k} + L_{34}^{s,k} + L_{43}^{s,k} + L_{44}^{s,k}) - (S_{33}^{s,k} + S_{34}^{s,k} + S_{43}^{s,k} + S_{44}^{s,k})^2 \bigg]$.
\end{widetext}

 The effective Onsager matrix elements can find all the thermoelectric coefficients. For the setup with the {BVT} probe with phase relaxation only, the thermal conductance is given by \cite{helical444, Whitney},
\begin{equation} \label{eq50}
\kappa'' = \frac{I_{2q}}{\Delta T}|_{I_{ic} = 0} =\frac{L''_{cV}L''_{qT} - L''_{cT}L''_{qV}}{L''_{cV}}.
\end{equation}
The Lorenz ratio for the setup with the {BVT} probe with phase relaxation only is then given by,
\begin{equation} \label{eq51}
W = \dfrac{\kappa^{''}}{\sigma''}, \text{where, } \sigma'' = L''_{cV}
\end{equation}
The Mathematica codes are available in GitHub ~\cite{s4}.

\section{Results}

\begin{figure*}
     \centering
     \begin{subfigure}[b]{0.40\textwidth}
         \centering
         \includegraphics[width=\textwidth]{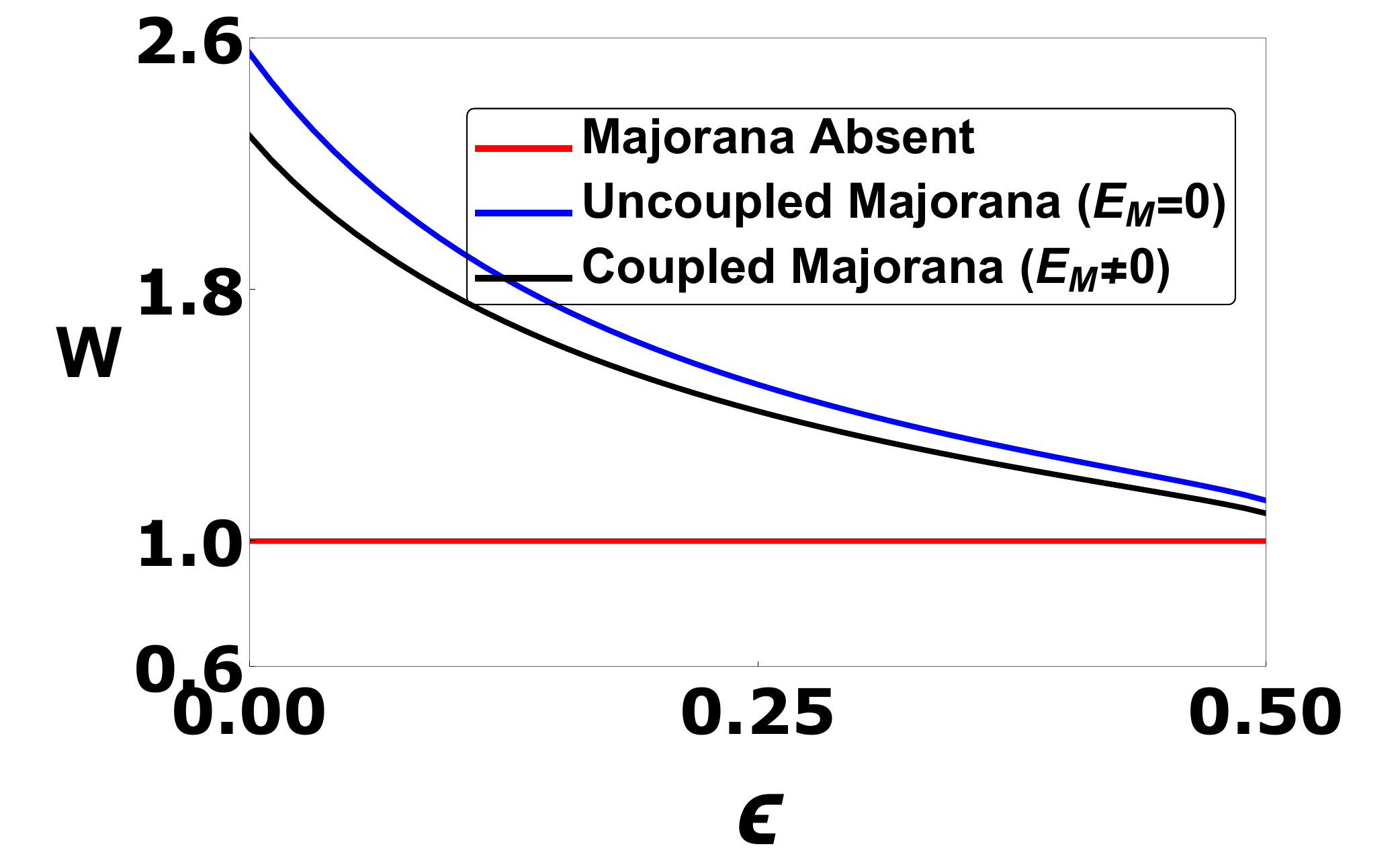}
         \caption{}
     \end{subfigure}
     \hspace{0.05cm}
     \begin{subfigure}[b]{0.40\textwidth}
         \centering
         \includegraphics[width=\textwidth]{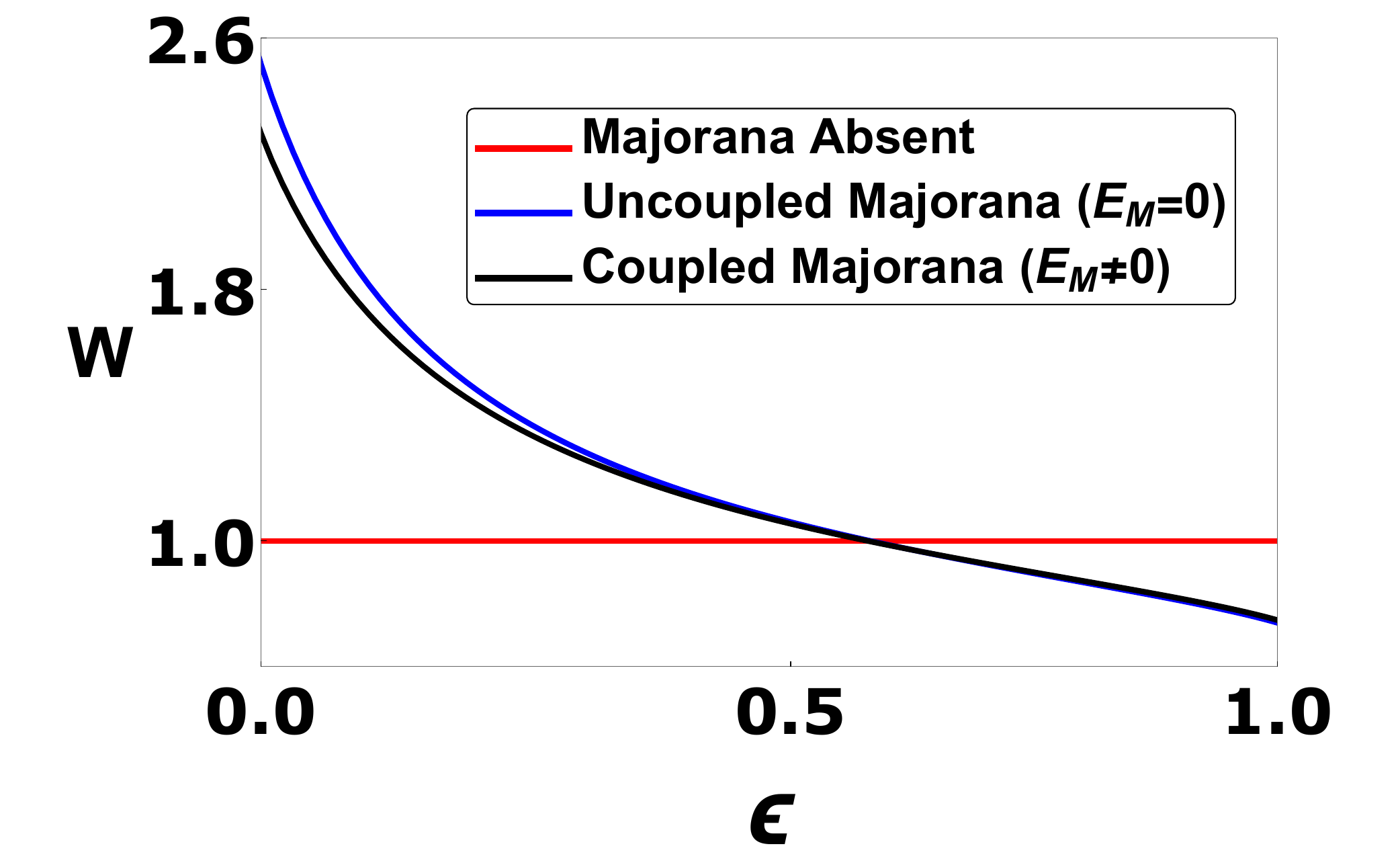}
         \caption{}
     \end{subfigure}
        \caption{{Effect of change in elastic scattering: Lorentz ratio ($W$) vs. the coupling strength ($\epsilon$) of the {BVT} probe for (a) with both phase and momentum relaxation (b) with phase relaxation. Parameters are $E_F = 10 \mu eV$, $E_M = 0.3 \mu eV$, $\Gamma = 1 \mu eV$, $\phi = \phi_0$ and elastic scattering at junction $J_1, J_2$, where $\phi = h c / e$ and $\alpha = \frac{4}{9}$ and represents elastic scattering at J1 and J2.}}
         \label{fig:2}
       \end{figure*}

In Fig. 2, we plot the Lorenz ratio vs. the strength of inelastic scattering for the setup with both phase and momentum relaxation \cite{buttikerboth}, and the setup with only phase relaxation \cite{PhysRevB.33.3020} in the presence of coupled MBS, individual MBS, and the absence of MBS. We use the parameters, Fermi energy $E_{F} = 10 \mu eV$, and flux $\phi = \phi_{0}$, $\phi_{0}$ being the flux quantum $hc/e$. In the absence of MBS, WF law is preserved in all three cases, i.e., $W = 1$ for the case of both phase and momentum relaxation (Fig. 2 (a)), and for case of only phase relaxation (Fig. 2 (b)) regardless of inelastic scattering $\epsilon$. For the presence of MBS, we consider two cases: individual MBS and coupled MBS. In our setup, we consider the MBS coupling strength $E_{M}  = 10 \mu eV$ when considering coupled MBS. The violation in Wiedemann-Franz law is highest when $E_{F}$ is close to $E_{M}$. Thus, to distinguish between individual MBS ($E_{M} = 0$), and coupled MBS ($E_{M} = 3 \mu eV$), we take $E_{F} = 10 \mu eV$.

For $\epsilon \to 0$, the {BVT} probe is completely disconnected, and there is no inelastic scattering in the setup for both cases. When MBS are uncoupled or individual ($E_{M} = 0$), the Lorenz ratio ($W$) is one at the limit of no inelastic scattering and decays with increasing strength of inelastic scattering ($\epsilon$). For the Luttinger liquid setup with individual MBS studied in Ref. \cite{buccheri}, the Lorenz ratio $W$ passes through one at $g = 1$, i.e., in the absence of inelastic scattering. Thus, our results for individual MBS corroborate the findings of the Luttinger liquid model studied in Ref. \cite{buccheri}. For case with both phase and momentum relaxation \cite{buttikerboth}, the Lorenz ratio scales with inelastic scattering ($\epsilon$) as {$W = \frac{0.76}{\epsilon^{1/4}}$}. Finally, for case with phase relaxation \cite{PhysRevB.33.3020}, the Lorenz ratio scales with $\epsilon$ as {$W = \frac{0.71}{\epsilon^{1/4}}$}. {Thus, for individual Majorana, the Lorenz ratio $W$ is inversely proportional to the strength of inelastic scattering $\epsilon$ for inelastic scattering with both phase and momentum relaxation with factor 1/4. Similarly,} the Lorenz ratio scales inversely with {$\epsilon^{1/4}$} for inelastic scattering with phase relaxation only.

       \begin{figure*}
     \centering
     \begin{subfigure}[b]{0.40\textwidth}
         \centering
         \includegraphics[width=\textwidth]{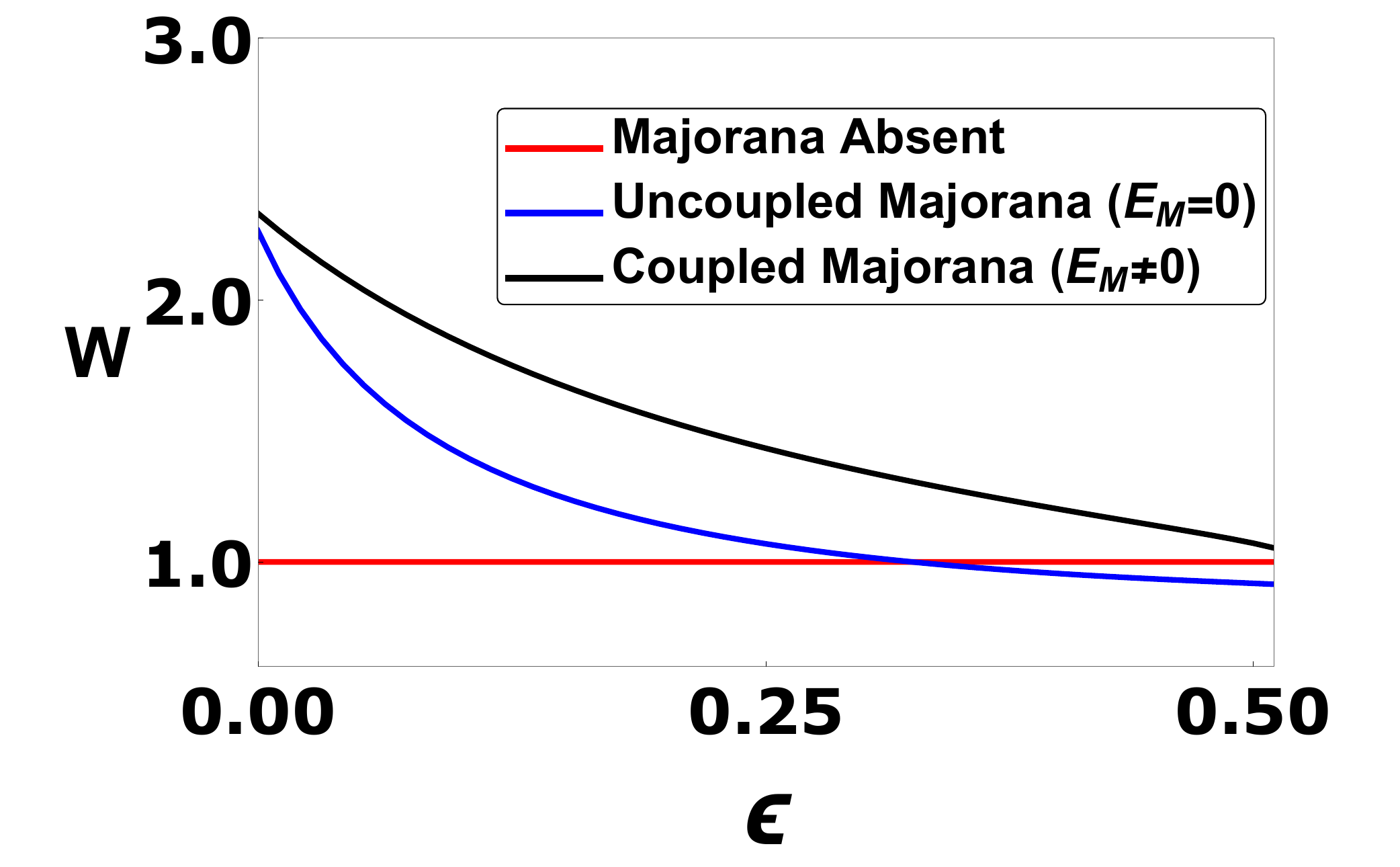}
         \caption{}
     \end{subfigure}
     \hspace{0.05cm}
     \begin{subfigure}[b]{0.40\textwidth}
         \centering
         \includegraphics[width=\textwidth]{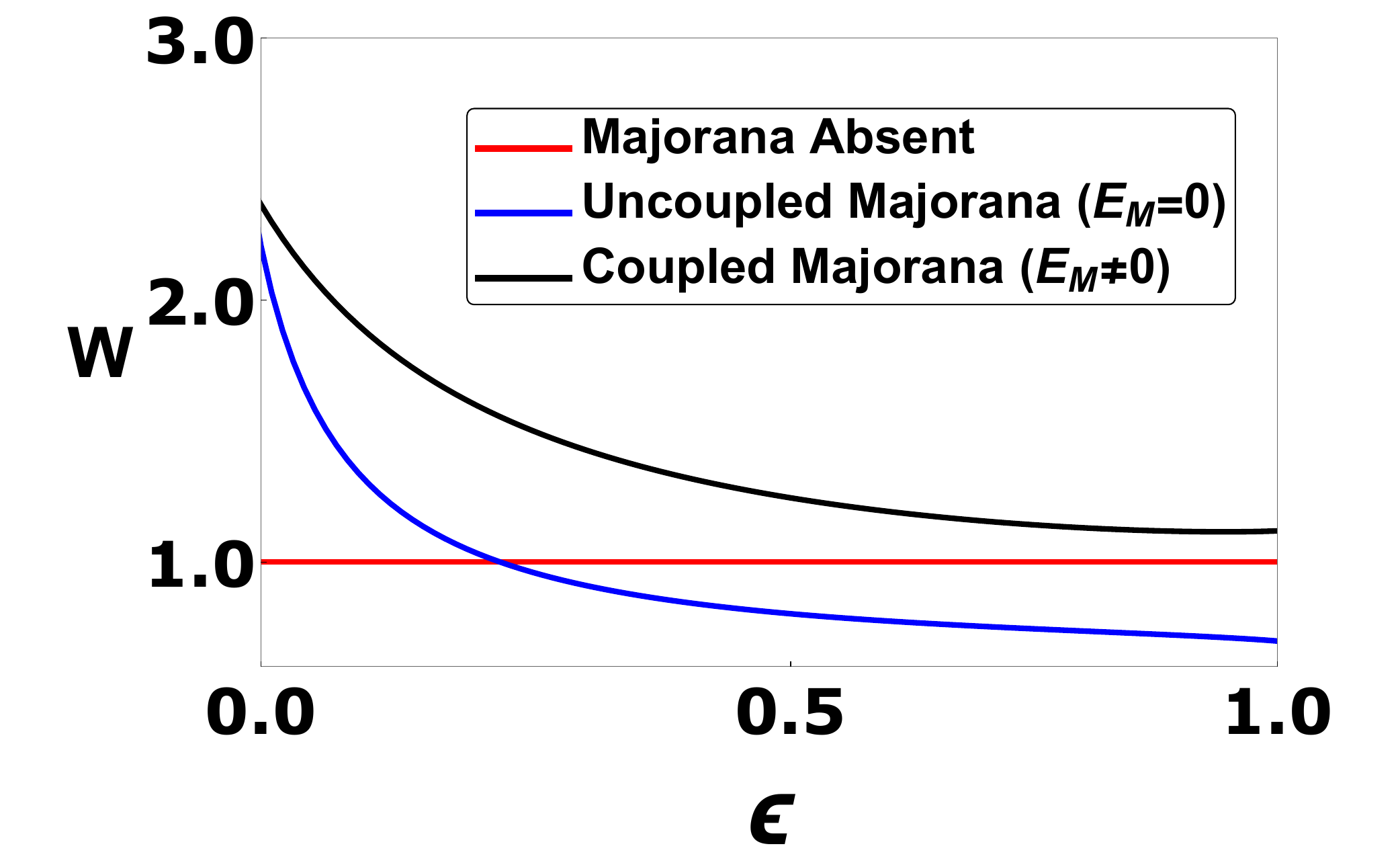}
         \caption{}
     \end{subfigure}
        \caption{{Effect of Majorana bound state energy: Lorentz ratio ($W$) vs. the coupling strength ($\epsilon$) of the {BVT} probe for (a) with both phase and momentum relaxation (b) with phase relaxation. Parameters are $E_F = 10 \mu eV$, $E_M = 2.55 \mu eV$, $\Gamma = 1 \mu eV$, $\phi = \phi_0$ and elastic scattering at junction $J_1, J_2$, where $\phi = h c / e$ and $\alpha = \frac{1}{2}$ and represents elastic scattering at J1 and J2.}}
         \label{fig:3}
       \end{figure*}

\begin{table*}[]
\resizebox{\textwidth}{!}{%
\begingroup

\setlength{\tabcolsep}{10pt} 
\renewcommand{\arraystretch}{1.5} 
\begin{tabular}{|l|l|l|}
\hline
Setup                                           & Coupled MBS & Individual MBS \\ \hline
Topological Kondo model \cite{buccheri}                        &    N/A         &  $W = \frac{0.66}{g}$              \\ \hline
\multicolumn{1}{|l|}{\begin{tabular}[c]{@{}l@{}}\scalebox{1.00}{{MBS ABI with both phase and momentum relaxation}}\\ \scalebox{1.00}{{{see Eq. (\ref{eq12}) and Ref. \cite{buttikerboth}}} }\end{tabular}} &    {$W \approx \frac{0.74}{\epsilon^{1/5}} $}         &    {$W \approx \frac{0.76}{\epsilon^{1/4}}$}            \\ \hline
  \multicolumn{1}{|l|}{\begin{tabular}[c]{@{}l@{}}\scalebox{1.00}{{MBS ABI with only phase relaxation}}\\ \scalebox{1.00}{{see Eq. (\ref{eq36}) and Ref. \cite{PhysRevB.33.3020}}}  \end{tabular}}          &   {$W \approx \frac{0.72}{\epsilon^{1/5}}  $}          &   {$W \approx \frac{0.71}{\epsilon^{1/4}}$}             \\ \hline
\end{tabular}
\endgroup
}

\caption{Comparing the violation of WF law in our setup with the topological Kondo model \cite{buccheri}. The authors of Ref. \cite{buccheri} only study individual MBS. We study both individual and coupled MBS. In all cases, we see power law dependence of the Lorenz ratio on inelastic scattering; the power law factors are different for inelastic scattering with both phase and momentum relaxation and with only phase relaxation.}
\end{table*}

For coupled MBS, $E_{M} = 0.3 \mu eV \neq 0$, we see that the Lorenz ratio is greater than one in the absence of inelastic scattering for all three cases, i.e. $W > 1$ for $\epsilon \to 0$ for case of both phase and momentum relaxation \cite{buttikerboth}, and $\epsilon \to 0$ for the case of only phase relaxation. In the presence of coupled MBS, PHS is broken such that the majority of electrons and holes travel in the same direction. It leads to a reduction in the net charge current and a commensurate increase in the heat current. Thus, in the presence of coupled MBS, WF law is violated even in the limit of zero inelastic scattering. The Lorenz ratio decays with increasing $\epsilon$ universally in the presence of MBS, regardless of whether they are coupled or uncoupled. Thus, in the presence of coupled MBS, inelastic scattering can recover the WF law at a particular $\epsilon$. From Fig. 2, we find that the Lorenz ratio $W$ scaling with respect to inelastic scattering $\epsilon$ also follows the power-law for coupled MBS. For case of both phase and momentum relaxation \cite{buttikerboth} (Fig. 2 (a)), the Lorenz ratio scales with inelastic scattering ($\epsilon$) as {$W = \frac{0.74}{\epsilon^{1/5}}$}. For the case of only phase relaxation \cite{PhysRevB.33.3020}, the Lorenz ratio scales with $\epsilon$ as {$W = \frac{0.72}{\epsilon^{1/5}}$}. Thus, the Lorenz ratio for coupled MBS is inversely proportional to $\epsilon^{1/5}$ when both phase and momentum relaxation are present. In the presence of inelastic scattering with phase relaxation only, the Lorenz ratio is again inversely proportional to $\epsilon^{1/5}$. 

\section{Analysis}
\subsection{{Comparison with the work of Buccheri et.al. \cite{buccheri}}}

We compare our results with the Luttinger liquid model studied in Ref. \cite{buccheri} in Table I. The setup in Ref. \cite{buccheri} considers a quantum dot junction hosting MBS in the topological Kondo regime using the many-body formalism. The Luttinger parameter $g$ describes electron-electron interaction in the many-body picture with $g = 1$ corresponding to no interaction. The authors in Ref. \cite{buccheri} report that weakly coupled MBS ($E_{M} \approx 0$) violate WF law. The authors show that the Lorenz ratio scales as $W = \frac{2}{3g}$, i.e., the Lorenz ratio is inversely proportional to the Luttinger parameter in the presence of uncoupled MBS. In {contrast in our setup, the Lorentz ratio is inversely proportional to $\epsilon^{1/4}$ for {elastic scattering with} both phase and momentum relaxation. Similarly, for {only} phase relaxation case, {Lorentz ratio} is inversely proportional to $\epsilon^{1/4}$}. {Table I shows the power law {scaling in} our setup, either with phase and
momentum relaxation or with only phase relaxation. In either case, one identical distinct power law violations for uncoupled and coupled MBS. This provides another unique method to detect MBS in mesoscopic superconducting junctions.} 

From Table I, we can see that the scaling of the Lorenz ratio changes when both phase and momentum relaxation are present in the setup and when only phase relaxation is present in the setup. In our setup, the major violations in WF law arise due to the breaking of PHS by the MBS in the upper arm. In the upper arm, MBS causes electron-hole mixing and backscattering \cite{nilssonMBS}. For both phase and momentum relaxation, the electrons and holes in the lower arm are transmitted, and backscattered \cite{inelasticWFLAW}. This backscattering can counteract the breaking of PHS by MBS and lead to a weaker violation. Similar effects have been observed in Ref. \cite{recovery} wherein the authors observe the restoration of WF law due to inelastic scattering. When only phase relaxation is present, the electrons and holes continue to travel in the same direction they initially traveled. Thus, they do not strongly counteract the breaking of PHS by the MBS. It can explain why the scatterer affects the violations more strongly with only phase relaxation. Thus, we see that the scaling changes in the presence of only phase relaxation. The unique scaling of the Lorenz ratio in the presence of MBS with only phase relaxation and with both phase and momentum relaxation can be used as a signature for MBS.

\subsection{Effect of elastic scattering at 3 way junctions J1 and J2}

{We look for the effects of elastic scattering on the power law scaling. In our calculation till now, we have considered strength of elastic scattering at the 3 way junctions J1 and J2 to be $\alpha = \frac{1}{2}$, see Eq. (\ref{eq4}). We can also study the power law scaling for a different value of elastic scattering $\alpha$ at junctions J1 and J2. We only have elastic scattering, as phase of electron is preserved in the scattering and terminals 1 and 2 are current probes. In Fig. \ref{fig:2}, we consider $\alpha = \frac{4}{9}$ and we observe that the power law scaling does not change. The reason for using this S-matrix as in Eq. (4) is that it corresponds exactly to a waveguide result, see Ref. \cite{cbook}. We observe that for the uncoupled MBS case, the Lorentz ratio with both phase and momentum relaxation follows a power law $\frac{0.81}{\epsilon^{1/4}}$, while only for phase relaxation, it follows a power law $\frac{0.80}{\epsilon^{1/4}}$, see Fig. \ref{fig:2}. Similarly, for the coupled MBS case, the Lorentz ratio for both phase and momentum relaxation followed a power law of $\frac{0.89}{\epsilon^{1/5}}$, where as for phase relaxation, it is $\frac{0.91}{\epsilon^{1/5}}$, see Fig. \ref{fig:2}. We observe that the presence of elastic scattering does affect the prefactor to some extent but the power law in the violation of the Wiedemann-Franz law remains same.}

\subsection{Effect of Majorana bound state energy}

{In Fig. \ref{fig:12}, we observe that the coupled MBS has weaker violation compared to the uncoupled MBS. However, this is not always the case. In Fig. 2, we had considered the Majorana bound state energy $E_M =  0.3 \mu eV$ for the coupled MBS. However, if we consider $E_M = 2.55 \mu eV$, then the coupled MBS can have stronger violation than uncoupled MBS ($E_M = 0$), see, Fig. \ref{fig:3}. We observe that while the Lorentz ratio for the phase and momentum relaxation follows a power law of $\frac{0.76}{\epsilon^{1/4}}$, while only for phase relaxation follows a power law of $\frac{0.71}{\epsilon^{1/4}}$ for the uncoupled MBS. For coupled MBS, the Lorentz ratio with phase and momentum relaxation follows a power law $\frac{1.05}{\epsilon^{1/5}}$, where as for only phase relaxation follows a power law $\frac{1.11}{\epsilon^{1/5}}$. Herein also although the power law remains unchanged, the numerical prefactor undergoes a significant change due to change in $E_M$ value.}

\section{Conclusion}
We studied the violation of Wiedemann-Franz law in the presence and absence of MBS and the scaling of the Lorenz ratio concerning the strength of inelastic scattering induced by a {BVT} probe. We find that WF law is only violated in the presence of MBS. We studied the scaling of the Lorenz ratio in the presence of inelastic scattering with both phase and momentum relaxation and with only phase relaxation and compared our results with those of the Luttinger liquid model studied in Ref. \cite{buccheri}. We showed that the Lorenz ratio decays with increasing strength of inelastic scattering when MBS are present, irrespective of whether they are individual or coupled. The Luttinger liquid model predicts that the Lorenz ratio is inversely proportional to the Luttinger parameter with $W(g) = \frac{2}{3g}$. In the presence of individual MBS with both phase and momentum relaxation, the Lorenz ratio scales with inelastic scattering ($\epsilon$) as {$W = \frac{0.76}{\epsilon^{1/4}}$}, whereas for phase relaxation only, the Lorenz ratio scales with $\epsilon$ as {$W = \frac{0.71}{\epsilon^{1/4}}$}. When MBS are coupled, the scaling is similar with {$W = \frac{0.74}{\epsilon^{1/5}}$} for phase and momentum relaxation, and {$W = \frac{0.72}{\epsilon^{1/5}}$} with phase relaxation only \cite{PhysRevB.33.3020}. {We also observe that the presence of elastic scattering at J1 and J2 and Majorana bound states energy does not have any impact on the power law, but the numerical prefactor significantly changes.} The results show that the electron-electron interaction in the Luttinger liquid model is similar to the inelastic scattering induced phenomenologically by the {BVT} probe \cite{dvira} with both phase and momentum relaxation \cite{buttikerboth, heikkila2013physics}. Further, we show that for coupled MBS, the Lorenz ratio is greater than one when inelastic scattering is absent and decays with increasing $\epsilon$. For individual MBS, the Lorenz ratio is conserved without inelastic scattering and decays when inelastic scattering is introduced. This distinct behavior of the Lorenz ratio in the scaling can be used to detect MBS.

{There can be several ways of implementing this work experimentally.
Refs. \cite{veldhorst2012josephson, sacepe2011gate} show the experimental realization of proximity-induced superconductivity in topological insulators, which is a key component of our setup. The superconductor-ferromagnet junction has also been experimentally realized before \cite{PhysRevB.69.224502}.
On the question on effect of magnetic field. The magnetic field is shielded such that it does not influence the magnetization of the ferromagnet. This is similar to the Aharonov-Bohm effect seen in normal metal rings and semiconductors. As such, the magnetization in the ferromagnets will be finite. Our setup may be realized in a HgTe quantum well, see \cite{nilssonMBS}. { The couplers in our study can represent either elastic scattering at J1, J2 or inelastic scattering at the {BVT} probe. In our approach, the use of a {BVT} probe plays a critical role. Experimentally, significant progress has already been made in deriving such {BVT} probes with the realization of small Ohmic contacts that function as voltage probes, as demonstrated in a pioneering study \cite{PhysRevLett.102.236802}. A voltage probe is a device through which no net charge current flows, achieved by applying a specific voltage bias. However, our proposal requires a {BVT} probe, which is slightly more advanced, wherein both the net charge and heat currents are zero. Similarly, the couplers J1 and J2 allow elastic scattering through them. The net charge and heat currents via these couplers are non-zero, in contrast to the other coupler through which the net charge and heat currents are zero. These couplers can also be modeled as a three-way junction, similar to the coupler used for inelastic scattering. These advancements in experimental realizations underscore the feasibility and importance of the {BVT} probe in understanding fundamental quantum transport phenomena in mesoscopic systems, which are important for considering many-body effects in mesoscopic samples phenomenologically \cite{dvira, kilgour2015charge, kilgour2015tunneling}.}

\appendix

\section{Expression for charge and heat current and Onsager coefficients with phase and momentum relaxation}

{Using Landauer-Buttiker scattering theory, we can find the charge and heat currents in each terminal of the setup described in Fig. 1 (a). In our work, the left and right terminals are at voltages $V_{1} = 0$, and $V_{2} = V$, respectively. {Similarly, the temperatures of the left and right terminals are $T_1$ and $T_1 + \Delta T_R$ respectively.} The {BVT} probe is at temperature $T + \Delta T$, and voltage $V_{3}$. From Eq. (\ref{eq11}), we can write the charge and heat currents for our setup in terms of the Onsager elements as,}

{
\begin{equation} \label{eq14}
{
\begin{pmatrix}
I^{\uparrow, e}_{1c}\\
I^{\uparrow, h}_{1c}\\
I^{\uparrow, e}_{2c}\\
I^{\uparrow, h}_{2c}\\
I^{\uparrow, e}_{3c}\\
I^{\uparrow, h}_{3c}\\
\end{pmatrix} = \begin{pmatrix}
L^{\uparrow;e}_{12;cV} & L^{\uparrow ;e}_{13;cV} & L^{\uparrow ;e}_{12;cT} & L^{\uparrow ;e}_{13;cT} \\
L^{\uparrow;h}_{12;cV} & L^{\uparrow;h}_{13;cV} & L^{\uparrow ;h}_{12;cT} & L^{\uparrow ;h}_{13;cT}\\
L^{\uparrow ;e}_{22;cV} & L^{\uparrow;e}_{23;cV} & L^{\uparrow ;e}_{23;cT} & L^{\uparrow ;e}_{23;cT} \\
L^{\uparrow ;h}_{22;cV} & L^{\uparrow ;h}_{23;cV} & L^{\uparrow;h}_{23;cT} & L^{\uparrow ;h}_{23;cT}\\
L^{\uparrow ;e}_{32;cV} & L^{\uparrow ;e}_{33;cV} & L^{\uparrow ;e}_{32;cT} & L^{\uparrow ;e}_{33;cT}  \\
L^{\uparrow ;h}_{32;cV} & L^{\uparrow ;h}_{33;cV} & L^{\uparrow ;h}_{32;cT} & L^{\uparrow ;h}_{33;cT}
\end{pmatrix} \begin{pmatrix}
V\\
V_{3}\\
\Delta T_R\\
{\Delta T}
\end{pmatrix},
}
\end{equation}
}

{
\begin{equation} \label{eq15}
{
\begin{pmatrix}
I^{\uparrow, e}_{1q}\\
I^{\uparrow, h}_{1q}\\
I^{\uparrow, e}_{2q}\\
I^{\uparrow, h}_{2q}\\
I^{\uparrow, e}_{3q}\\
I^{\uparrow, h}_{3q}\\
\end{pmatrix} = \begin{pmatrix}
L^{\uparrow;e}_{12;qV} & L^{\uparrow ;e}_{13;qV} & L^{\uparrow ;e}_{12;qT} & L^{\uparrow ;e}_{13;qT} \\
L^{\uparrow;h}_{12;qV} & L^{\uparrow;h}_{13;qV} & L^{\uparrow ;h}_{12;qT} & L^{\uparrow ;h}_{13;qT}\\
L^{\uparrow ;e}_{22;qV} & L^{\uparrow;e}_{23;qV} & L^{\uparrow ;e}_{23;qT} & L^{\uparrow ;e}_{23;qT} \\
L^{\uparrow ;h}_{22;qV} & L^{\uparrow ;h}_{23;qV} & L^{\uparrow;h}_{23;qT} & L^{\uparrow ;h}_{23;qT}\\
L^{\uparrow ;e}_{32;qV} & L^{\uparrow ;e}_{33;qV} & L^{\uparrow ;e}_{32;qT} & L^{\uparrow ;e}_{33;qT}  \\
L^{\uparrow ;h}_{32;qV} & L^{\uparrow ;h}_{33;qV} & L^{\uparrow ;h}_{32;qT} & L^{\uparrow ;h}_{33;qT}
\end{pmatrix} \begin{pmatrix}
V\\
V_{3}\\
\Delta T_R\\
{\Delta T}
\end{pmatrix},
}
\end{equation}}

{
\begin{equation} \label{eq16}
{
\begin{pmatrix}
I^{\downarrow, e}_{1c}\\
I^{\downarrow, h}_{1c}\\
I^{\downarrow, e}_{2c}\\
I^{\downarrow, h}_{2c}\\
I^{\downarrow, e}_{3c}\\
I^{\downarrow, h}_{3c}\\
\end{pmatrix} = \begin{pmatrix}
L^{\downarrow;e}_{12;cV} & L^{\downarrow ;e}_{13;cV} & L^{\downarrow ;e}_{12;cT} & L^{\downarrow ;e}_{13;cT} \\
L^{\downarrow;h}_{12;cV} & L^{\downarrow;h}_{13;cV} & L^{\downarrow ;h}_{12;cT} & L^{\downarrow ;h}_{13;cT}\\
L^{\downarrow ;e}_{22;cV} & L^{\downarrow;e}_{23;cV} & L^{\downarrow ;e}_{23;cT} & L^{\downarrow ;e}_{23;cT} \\
L^{\downarrow ;h}_{22;cV} & L^{\downarrow ;h}_{23;cV} & L^{\downarrow;h}_{23;cT} & L^{\downarrow ;h}_{23;cT}\\
L^{\downarrow ;e}_{32;cV} & L^{\downarrow ;e}_{33;cV} & L^{\downarrow ;e}_{32;cT} & L^{\downarrow ;e}_{33;cT}  \\
L^{\downarrow ;h}_{32;cV} & L^{\downarrow ;h}_{33;cV} & L^{\downarrow ;h}_{32;cT} & L^{\downarrow ;h}_{33;cT}
\end{pmatrix} \begin{pmatrix}
V\\
V_{3}\\
\Delta T_R\\
{\Delta T}
\end{pmatrix},
}
\end{equation}}

{
\begin{equation} \label{eq17}
{
\begin{pmatrix}
I^{\downarrow, e}_{1q}\\
I^{\downarrow, h}_{1q}\\
I^{\downarrow, e}_{2q}\\
I^{\downarrow, h}_{2q}\\
I^{\downarrow, e}_{3q}\\
I^{\downarrow, h}_{3q}\\
\end{pmatrix} = \begin{pmatrix}
L^{\downarrow;e}_{12;qV} & L^{\downarrow ;e}_{13;qV} & L^{\downarrow ;e}_{12;qT} & L^{\downarrow ;e}_{13;qT} \\
L^{\downarrow;h}_{12;qV} & L^{\downarrow;h}_{13;qV} & L^{\downarrow ;h}_{12;qT} & L^{\downarrow ;h}_{13;qT}\\
L^{\downarrow ;e}_{22;qV} & L^{\downarrow;e}_{23;qV} & L^{\downarrow ;e}_{23;qT} & L^{\downarrow ;e}_{23;qT} \\
L^{\downarrow ;h}_{22;qV} & L^{\downarrow ;h}_{23;qV} & L^{\downarrow;h}_{23;qT} & L^{\downarrow ;h}_{23;qT}\\
L^{\downarrow ;e}_{32;qV} & L^{\downarrow ;e}_{33;qV} & L^{\downarrow ;e}_{32;qT} & L^{\downarrow ;e}_{33;qT}  \\
L^{\downarrow ;h}_{32;qV} & L^{\downarrow ;h}_{33;qV} & L^{\downarrow ;h}_{32;qT} & L^{\downarrow ;h}_{33;qT}
\end{pmatrix} \begin{pmatrix}
V\\
V_{3}\\
\Delta T_R\\
{\Delta T}
\end{pmatrix},
}
\end{equation}}

{We are solving Eqs. (\ref{eq3}-\ref{eq8}) and using Eqs. (\ref{eq12}, \ref{eq14}) for the S-matrix of the {BVT} probe, one can find the transmission probabilities {$\mathcal{T}^{ss;kk}_{ij}$} and {$\mathcal{T}^{rs;lk}_{ij}$} in the setup. Plugging the transmission probabilities in Eqs. (\ref{eq14}-\ref{eq17}), along with Eq. (\ref{eq9}), allows us to calculate the charge and heat currents in each terminal. We represent the charge and heat currents in terms of the voltage and temperature biases in a matrix notation below:
The charge currents in each terminal due to spin-up electrons and holes are given by,}

{
\begin{equation} \label{eq18}
{
\begin{pmatrix}
I^{\uparrow, e}_{1c}\\
I^{\uparrow, h}_{1c}\\
I^{\uparrow, e}_{2c}\\
I^{\uparrow, h}_{2c}\\
I^{\uparrow, e}_{3c}\\
I^{\uparrow, h}_{3c}\\
\end{pmatrix} = \begin{pmatrix}
G(\mathcal{T}^{\uparrow; e}_{12}) & G(\mathcal{T}^{\uparrow ; e}_{13})  & S(\mathcal{T}^{\uparrow ; e}_{12}) & S(\mathcal{T}^{\uparrow ; e}_{13}) \\
G(\mathcal{T}^{\uparrow ; h}_{12}) & G(\mathcal{T}^{\uparrow ; h}_{13}) & S(\mathcal{T}^{\uparrow ; h}_{12}) & S(\mathcal{T}^{\uparrow ; h}_{13})\\
G(\mathcal{T}^{ \uparrow; e}_{22}) & G(\mathcal{T}^{\uparrow ; e}_{23}) & S(\mathcal{T}^{\uparrow; e}_{23}) & S(\mathcal{T}^{\uparrow; e}_{23})\\
G(\mathcal{T}^{\uparrow ; h}_{22}) & G(\mathcal{T}^{\uparrow ; h}_{23}) & S(\mathcal{T}^{\uparrow ; h}_{22}) & S(\mathcal{T}^{\uparrow ; h}_{23})\\
G(\mathcal{T}^{\uparrow; e}_{32}) & G(\mathcal{T}^{\uparrow; e}_{32}) & S(\mathcal{T}^{ \uparrow; ee}_{32}) & S(\mathcal{T}^{\uparrow; e}_{33})\\
G(\mathcal{T}^{\uparrow; h}_{32}) & G(\mathcal{T}^{\uparrow; h}_{3}) & S(\mathcal{T}^{\uparrow; h}_{32}) & S(\mathcal{T}^{\uparrow; h}_{33})\\
\end{pmatrix} \begin{pmatrix}
V\\
V_{3}\\
\Delta T_R\\
\Delta T
\end{pmatrix}},
\end{equation}}
{The corresponding heat currents due to spin-up electrons and holes are given by:}
{
\begin{equation} \label{eq19}
{
\begin{pmatrix}
I^{\uparrow, e}_{1q}\\
I^{\uparrow, h}_{1q}\\
I^{\uparrow, e}_{2q}\\
I^{\uparrow, h}_{2q}\\
I^{\uparrow, e}_{3q}\\
I^{\uparrow, h}_{3q}\\
\end{pmatrix} = \begin{pmatrix}
T S(\mathcal{T}^{\uparrow; e}_{12}) & T S(\mathcal{T}^{\uparrow ; e}_{13})  & L(\mathcal{T}^{\uparrow ; e}_{12}) & L(\mathcal{T}^{\uparrow ; e}_{13}) \\
T S(\mathcal{T}^{\uparrow ; h}_{12}) & T S(\mathcal{T}^{\uparrow ; h}_{13}) & L(\mathcal{T}^{\uparrow ; h}_{12}) & L(\mathcal{T}^{\uparrow ; h}_{13})\\
T S(\mathcal{T}^{ \uparrow; e}_{22}) & T S(\mathcal{T}^{\uparrow ; e}_{23}) & L(\mathcal{T}^{\uparrow; e}_{23}) & L(\mathcal{T}^{\uparrow; e}_{23})\\
T S(\mathcal{T}^{\uparrow ; h}_{22}) & T S(\mathcal{T}^{\uparrow ; h}_{23}) & L(\mathcal{T}^{\uparrow ; h}_{22}) & L(\mathcal{T}^{\uparrow ; h}_{23})\\
T S(\mathcal{T}^{\uparrow; e}_{32}) & T S(\mathcal{T}^{\uparrow; e}_{32}) & L(\mathcal{T}^{ \uparrow; ee}_{32}) & L(\mathcal{T}^{\uparrow; e}_{33})\\
T S(\mathcal{T}^{\uparrow; h}_{32}) & T S(\mathcal{T}^{\uparrow; h}_{3}) & L(\mathcal{T}^{\uparrow; h}_{32}) & L(\mathcal{T}^{\uparrow; h}_{33})\\
\end{pmatrix} \begin{pmatrix}
V\\
V_{3}\\
\Delta T_R\\
\Delta T
\end{pmatrix}},
\end{equation}}
{Similarly, for the spin-down electrons and holes, the charge currents are given as,}

{
\begin{equation} \label{eq20}
{
\begin{pmatrix}
I^{\downarrow, e}_{1c}\\
I^{\downarrow, h}_{1c}\\
I^{\downarrow, e}_{2c}\\
I^{\downarrow, h}_{2c}\\
I^{\downarrow, e}_{3c}\\
I^{\downarrow, h}_{3c}\\
\end{pmatrix} = \begin{pmatrix}
G(\mathcal{T}^{\downarrow; e}_{12}) & G(\mathcal{T}^{\downarrow ; e}_{13})  & S(\mathcal{T}^{\downarrow ; e}_{12}) & S(\mathcal{T}^{\downarrow ; e}_{13}) \\
G(\mathcal{T}^{\downarrow ; h}_{12}) & G(\mathcal{T}^{\downarrow ; h}_{13}) & S(\mathcal{T}^{\downarrow ; h}_{12}) & S(\mathcal{T}^{\downarrow ; h}_{13})\\
G(\mathcal{T}^{ \downarrow; e}_{22}) & G(\mathcal{T}^{\downarrow ; e}_{23}) & S(\mathcal{T}^{\downarrow; e}_{23}) & S(\mathcal{T}^{\downarrow; e}_{23})\\
G(\mathcal{T}^{\downarrow ; h}_{22}) & G(\mathcal{T}^{\downarrow ; h}_{23}) & S(\mathcal{T}^{\downarrow ; h}_{22}) & S(\mathcal{T}^{\downarrow ; h}_{23})\\
G(\mathcal{T}^{\downarrow; e}_{32}) & G(\mathcal{T}^{\downarrow; e}_{32}) & S(\mathcal{T}^{ \downarrow; ee}_{32}) & S(\mathcal{T}^{\downarrow; e}_{33})\\
G(\mathcal{T}^{\downarrow; h}_{32}) & G(\mathcal{T}^{\downarrow; h}_{3}) & S(\mathcal{T}^{\downarrow; h}_{32}) & S(\mathcal{T}^{\downarrow; h}_{33})\\
\end{pmatrix} \begin{pmatrix}
V\\
V_{3}\\
\Delta T_R\\
\Delta T
\end{pmatrix}},
\end{equation}}
{and the corresponding heat currents due to spin-down electrons and holes are given by:}
{
\begin{equation} \label{eq21}
{
\begin{pmatrix}
I^{\downarrow, e}_{1q}\\
I^{\downarrow, h}_{1q}\\
I^{\downarrow, e}_{2q}\\
I^{\downarrow, h}_{2q}\\
I^{\downarrow, e}_{3q}\\
I^{\downarrow, h}_{3q}\\
\end{pmatrix} = \begin{pmatrix}
T S(\mathcal{T}^{\downarrow; e}_{12}) & T S(\mathcal{T}^{\downarrow ; e}_{13})  & L(\mathcal{T}^{\downarrow ; e}_{12}) & L(\mathcal{T}^{\downarrow ; e}_{13}) \\
T S(\mathcal{T}^{\downarrow ; h}_{12}) & T S(\mathcal{T}^{\downarrow ; h}_{13}) & L(\mathcal{T}^{\downarrow ; h}_{12}) & L(\mathcal{T}^{\downarrow ; h}_{13})\\
T S(\mathcal{T}^{ \downarrow; e}_{22}) & T S(\mathcal{T}^{\downarrow ; e}_{23}) & L(\mathcal{T}^{\downarrow; e}_{23}) & L(\mathcal{T}^{\downarrow; e}_{23})\\
T S(\mathcal{T}^{\downarrow ; h}_{22}) & T S(\mathcal{T}^{\downarrow ; h}_{23}) & L(\mathcal{T}^{\downarrow ; h}_{22}) & L(\mathcal{T}^{\downarrow ; h}_{23})\\
T S(\mathcal{T}^{\downarrow; e}_{32}) & T S(\mathcal{T}^{\downarrow; e}_{32}) & L(\mathcal{T}^{ \downarrow; ee}_{32}) & L(\mathcal{T}^{\downarrow; e}_{33})\\
T S(\mathcal{T}^{\downarrow; h}_{32}) & T S(\mathcal{T}^{\downarrow; h}_{3}) & L(\mathcal{T}^{\downarrow; h}_{32}) & L(\mathcal{T}^{\downarrow; h}_{33})\\
\end{pmatrix} \begin{pmatrix}
V\\
V_{3}\\
\Delta T_R\\
\Delta T
\end{pmatrix}},
\end{equation}}

{where $G(X)$ is the electrical conductance for particles with transmission probability $X$, $S(X)$ is the Seebeck coefficient for particles with transmission probability $X$, and $L(X)$ is the thermal conductance for particles with transmission probability $X$.  We can calculate the charge and heat currents for phase and momentum relaxation given in Eq. (\ref{eq12}) by using the respective S-matrix for the {BVT} probe \cite{ dvira} when solving for the transmission probabilities {$\mathcal{T}^{ss;kk}_{ij}$ and $\mathcal{T}^{rs;lk}_{ij}$}. In Eqs. (\ref{eq18}-\ref{eq21}) the electrical conductance {$G(\mathcal{T}^{s;k}_{ij}(E))$} is given by,}
{
\begin{equation} \label{eq22}
\begin{split}
{G(\mathcal{T}^{s;k}_{ij}(E)) = G_{0}\int_{-\infty}^{\infty}dE \left(-\frac{df}{dE}\right)}\\ {(\delta_{ij} - \mathcal{T}^{ss;k}_{ij}(E) + \mathcal{T}^{rs;lk}_{ij}(E))}
\end{split}
\end{equation}}
{the Seebeck coefficient $S(\mathcal{T}^{s;k}_{ij}(E))$ is given by,}
{
\begin{equation} \label{eq23}
\begin{split}
{S(\mathcal{T}^{s;k}_{ij}(E)) = G_{0}\int_{-\infty}^{\infty}dE (E -\mu) \left(-\frac{df}{dE}\right)}\\ {(\delta_{ij} - \mathcal{T}^{ss;k}_{ij}(E) + \mathcal{T}^{rs;lk}_{ij}(E))}
\end{split}
\end{equation}}
{The thermal conductance $L(\mathcal{T}^{s;k}_{ij}(E))$ is given by,}
{
\begin{equation} \label{eq24}
\begin{split}
{S(\mathcal{T}^{s;k}_{ij}(E)) = G_{0}\int_{-\infty}^{\infty}dE (E -\mu)^2 \left(-\frac{df}{dE}\right)}\\ {(\delta_{ij} - \mathcal{T}^{ss;k}_{ij}(E) + \mathcal{T}^{rs;lk}_{ij}(E))}
\end{split}
\end{equation}}

\section{Expression for charge and heat current and Onsager coefficients with phase relaxation}
{Similar to the phase and momentum relaxation case, we can relate the charge and heat currents to the voltage and temperature difference using Landauer-Buttiker scattering theory. We set the voltage of the left terminal at zero and the temperature at $T$, while the right terminal is at voltage $V$ and temperature $T + \Delta T$. Using Eq. (\ref{eq11}), we relate the charge and heat currents to the voltage and temperature biases in matrix form as,}
{
\begin{equation} \label{eq37}
\resizebox{1\hsize}{!}{%
$
{\begin{pmatrix}
I^{\uparrow, e}_{1c}\\
I^{\uparrow, h}_{1c}\\
I^{\uparrow, e}_{2c}\\
I^{\uparrow, h}_{2c}\\
I^{\uparrow, e}_{3c}\\
I^{\uparrow, h}_{3c}\\
I^{\uparrow, e}_{4c}\\
I^{\uparrow, h}_{4c}\\
\end{pmatrix} = \begin{pmatrix}
L^{\uparrow;e}_{12;cV} & L^{ \uparrow;e}_{13;cV} & L^{\uparrow;e}_{14;cV} & L^{ \uparrow;e}_{12;cT} & L^{ \uparrow;e}_{13;cT} & L^{\uparrow;e}_{14;cT} \\
L^{\uparrow;h}_{12;cV} & L^{\uparrow;h}_{13;cV} &L^{\uparrow;h}_{14;cV} & L^{ \uparrow;h}_{12;cT} & L^{\uparrow;h}_{13;cT} & L^{ \uparrow;h}_{14;cT}\\
L^{\uparrow;e}_{22;cV} & L^{\uparrow;e}_{23;cV} & L^{\uparrow ;e}_{24;cV} & L^{ \uparrow;e}_{22;cT} & L^{\uparrow;e}_{23;cT} & L^{\uparrow;e}_{24;cT} \\
L^{\uparrow;h}_{22;cV} & L^{\uparrow;h}_{23;cV} &L^{\uparrow;h}_{24;cV} & L^{\uparrow;h}_{22;cT} & L^{\uparrow;h}_{23;cT} & L^{\uparrow;e}_{24;cT}\\
L^{\uparrow;e}_{32;cV} & L^{\uparrow;e}_{33;cV} & L^{ \uparrow;e}_{34;cV} & L^{\uparrow;e}_{32;cT} & L^{\uparrow;e}_{33;cT} & L^{\uparrow;e}_{34;cT} \\
L^{\uparrow;h}_{32;cV} & L^{ \uparrow;h}_{33;cV} &L^{ \uparrow;h}_{34;cV} & L^{\uparrow;h}_{32;cT} & L^{\uparrow;h}_{33;cT} & L^{\uparrow;h}_{34;cT} \\
L^{\uparrow;e}_{42;cV} & L^{\uparrow;e}_{43;cV} & L^{\uparrow;e}_{44;cV} & L^{\uparrow \uparrow;e}_{42;cT} & L^{\uparrow;e}_{43;cT} & L^{\uparrow;e}_{44;cT} \\
L^{\uparrow;h}_{42;cV} & L^{\uparrow;h}_{43;cV} &L^{\uparrow;h}_{44;cV} & L^{\uparrow;h}_{42;cT} & L^{\uparrow;h}_{43;cT} & L^{\uparrow;h}_{44;cT}
\end{pmatrix} \begin{pmatrix}
V\\
V_{3}\\
V_{3}\\
\Delta T_R\\
\Delta T\\
\Delta T
\end{pmatrix}},
$}
\end{equation}}
{
\begin{equation} \label{eq38}
\resizebox{1\hsize}{!}{%
$
{\begin{pmatrix}
I^{\uparrow, e}_{1q}\\
I^{\uparrow, h}_{1q}\\
I^{\uparrow, e}_{2q}\\
I^{\uparrow, h}_{2q}\\
I^{\uparrow, e}_{3q}\\
I^{\uparrow, h}_{3q}\\
I^{\uparrow, e}_{4q}\\
I^{\uparrow, h}_{4q}\\
\end{pmatrix} = \begin{pmatrix}
L^{\uparrow;e}_{12;qV} & L^{ \uparrow;e}_{13;qV} & L^{\uparrow;e}_{14;qV} & L^{ \uparrow;e}_{12;qT} & L^{ \uparrow;e}_{13;qT} & L^{\uparrow;e}_{14;qT} \\
L^{\uparrow;h}_{12;qV} & L^{\uparrow;h}_{13;qV} &L^{\uparrow;h}_{14;qV} & L^{ \uparrow;h}_{12;qT} & L^{\uparrow;h}_{13;qT} & L^{ \uparrow;h}_{14;qT}\\
L^{\uparrow;e}_{22;qV} & L^{\uparrow;e}_{23;qV} & L^{\uparrow ;e}_{24;qV} & L^{ \uparrow;e}_{22;qT} & L^{\uparrow;e}_{23;qT} & L^{\uparrow;e}_{24;qT} \\
L^{\uparrow;h}_{22;qV} & L^{\uparrow;h}_{23;qV} &L^{\uparrow;h}_{24;qV} & L^{\uparrow;h}_{22;qT} & L^{\uparrow;h}_{23;qT} & L^{\uparrow;e}_{24;qT}\\
L^{\uparrow;e}_{32;qV} & L^{\uparrow;e}_{33;qV} & L^{ \uparrow;e}_{34;qV} & L^{\uparrow;e}_{32;qT} & L^{\uparrow;e}_{33;qT} & L^{\uparrow;e}_{34;qT} \\
L^{\uparrow;h}_{32;qV} & L^{ \uparrow;h}_{33;qV} &L^{ \uparrow;h}_{34;qV} & L^{\uparrow;h}_{32;qT} & L^{\uparrow;h}_{33;qT} & L^{\uparrow;h}_{34;qT} \\
L^{\uparrow;e}_{42;qV} & L^{\uparrow;e}_{43;qV} & L^{\uparrow;e}_{44;qV} & L^{\uparrow \uparrow;e}_{42;qT} & L^{\uparrow;e}_{43;qT} & L^{\uparrow;e}_{44;qT} \\
L^{\uparrow;h}_{42;qV} & L^{\uparrow;h}_{43;qV} &L^{\uparrow;h}_{44;qV} & L^{\uparrow;h}_{42;qT} & L^{\uparrow;h}_{43;qT} & L^{\uparrow;h}_{44;qT}
\end{pmatrix} \begin{pmatrix}
V\\
V_{3}\\
V_{3}\\
\Delta T_R\\
\Delta T\\
\Delta T
\end{pmatrix}},
$}
\end{equation}}

{
\begin{equation} \label{eq39}
\resizebox{1\hsize}{!}{%
$
{\begin{pmatrix}
I^{\downarrow, e}_{1c}\\
I^{\downarrow, h}_{1c}\\
I^{\downarrow, e}_{2c}\\
I^{\downarrow, h}_{2c}\\
I^{\downarrow, e}_{3c}\\
I^{\downarrow, h}_{3c}\\
I^{\downarrow, e}_{4c}\\
I^{\downarrow, h}_{4c}\\
\end{pmatrix} = \begin{pmatrix}
L^{\downarrow;e}_{12;cV} & L^{ \downarrow;e}_{13;cV} & L^{\downarrow;e}_{14;cV} & L^{ \downarrow;e}_{12;cT} & L^{ \downarrow;e}_{13;cT} & L^{\downarrow;e}_{14;cT} \\
L^{\downarrow;h}_{12;cV} & L^{\downarrow;h}_{13;cV} &L^{\downarrow;h}_{14;cV} & L^{ \downarrow;h}_{12;cT} & L^{\downarrow;h}_{13;cT} & L^{ \downarrow;h}_{14;cT}\\
L^{\downarrow;e}_{22;cV} & L^{\downarrow;e}_{23;cV} & L^{\downarrow ;e}_{24;cV} & L^{ \downarrow;e}_{22;cT} & L^{\downarrow;e}_{23;cT} & L^{\downarrow;e}_{24;cT} \\
L^{\downarrow;h}_{22;cV} & L^{\downarrow;h}_{23;cV} &L^{\downarrow;h}_{24;cV} & L^{\downarrow;h}_{22;cT} & L^{\downarrow;h}_{23;cT} & L^{\downarrow;e}_{24;cT}\\
L^{\downarrow;e}_{32;cV} & L^{\downarrow;e}_{33;cV} & L^{ \downarrow;e}_{34;cV} & L^{\downarrow;e}_{32;cT} & L^{\downarrow;e}_{33;cT} & L^{\downarrow;e}_{34;cT} \\
L^{\downarrow;h}_{32;cV} & L^{ \downarrow;he}_{33;cV} &L^{ \downarrow;he}_{34;cV} & L^{\downarrow;h}_{32;cT} & L^{\downarrow;h}_{33;cT} & L^{\downarrow;h}_{34;cT} \\
L^{\downarrow;e}_{42;cV} & L^{\downarrow;e}_{43;cV} & L^{\downarrow;e}_{44;cV} & L^{\downarrow;e}_{42;cT} & L^{\downarrow;e}_{43;cT} & L^{\downarrow;e}_{44;cT} \\
L^{\downarrow;h}_{42;cV} & L^{\downarrow;h}_{43;cV} & L^{\downarrow;h}_{44;cV} & L^{\downarrow;h}_{42;cT} & L^{\downarrow;h}_{43;cT} & L^{\downarrow;h}_{44;cT}
\end{pmatrix} \begin{pmatrix}
V\\
V_{3}\\
V_{3}\\
\Delta T_R\\
\Delta T\\
\Delta T
\end{pmatrix}},
$}
\end{equation}}

{
\begin{equation} \label{eq40}
\resizebox{1\hsize}{!}{%
$
{\begin{pmatrix}
I^{\downarrow, e}_{1q}\\
I^{\downarrow, h}_{1q}\\
I^{\downarrow, e}_{2q}\\
I^{\downarrow, h}_{2q}\\
I^{\downarrow, e}_{3q}\\
I^{\downarrow, h}_{3q}\\
I^{\downarrow, e}_{4q}\\
I^{\downarrow, h}_{4q}\\
\end{pmatrix} = \begin{pmatrix}
L^{\downarrow;e}_{12;qV} & L^{ \downarrow;e}_{13;qV} & L^{\downarrow;e}_{14;qV} & L^{ \downarrow;e}_{12;qT} & L^{ \downarrow;e}_{13;qT} & L^{\downarrow;e}_{14;qT} \\
L^{\downarrow;h}_{12;qV} & L^{\downarrow;h}_{13;qV} &L^{\downarrow;h}_{14;qV} & L^{ \downarrow;h}_{12;qT} & L^{\downarrow;h}_{13;qT} & L^{ \downarrow;h}_{14;qT}\\
L^{\downarrow;e}_{22;qV} & L^{\downarrow;e}_{23;qV} & L^{\downarrow ;e}_{24;qV} & L^{ \downarrow;e}_{22;qT} & L^{\downarrow;e}_{23;qT} & L^{\downarrow;e}_{24;qT} \\
L^{\downarrow;h}_{22;qV} & L^{\downarrow;h}_{23;qV} &L^{\downarrow;h}_{24;qV} & L^{\downarrow;h}_{22;qT} & L^{\downarrow;h}_{23;qT} & L^{\downarrow;e}_{24;qT}\\
L^{\downarrow;e}_{32;qV} & L^{\downarrow;e}_{33;qV} & L^{ \downarrow;e}_{34;qV} & L^{\downarrow;e}_{32;qT} & L^{\downarrow;e}_{33;qT} & L^{\downarrow;e}_{34;qT} \\
L^{\downarrow;h}_{32;qV} & L^{ \downarrow;he}_{33;qV} &L^{ \downarrow;he}_{34;qV} & L^{\downarrow;h}_{32;qT} & L^{\downarrow;h}_{33;qT} & L^{\downarrow;h}_{34;qT} \\
L^{\downarrow;e}_{42;qV} & L^{\downarrow;e}_{43;qV} & L^{\downarrow;e}_{44;qV} & L^{\downarrow;e}_{42;qT} & L^{\downarrow;e}_{43;qT} & L^{\downarrow;e}_{44;qT} \\
L^{\downarrow;h}_{42;qV} & L^{\downarrow;h}_{43;qV} & L^{\downarrow;h}_{44;qV} & L^{\downarrow;h}_{42;qT} & L^{\downarrow;h}_{43;qT} & L^{\downarrow;h}_{44;qT}
\end{pmatrix} \begin{pmatrix}
V\\
V_{3}\\
V_{3}\\
\Delta T_R\\
\Delta T\\
\Delta T
\end{pmatrix}},
$}
\end{equation}}

{Finding the transmission probabilities {$\mathcal{T}^{ss;kk}_{ij}$ and $\mathcal{T}^{rs;lk}_{ij}$} from Eqs. (\ref{eq3}-\ref{eq8}) and using Eq. (\ref{eq36}) for the S-matrix of the inelastic scatterer, we can find the charge current in each terminal of the setup in Fig. 1 (b), i.e., with only phase relaxation as,}

\begin{widetext}
{\begin{equation} \label{eq41}
{\begin{pmatrix}
I_{1c}^{\uparrow, e}\\
I_{1c}^{\uparrow, h}\\
I_{2c}^{\uparrow, e}\\
I_{2c}^{\uparrow, h}\\
I_{3c}^{\uparrow, e}\\
I_{3c}^{\uparrow, h}\\
I_{4c}^{\uparrow, e}\\
I_{4c}^{\uparrow, h}
\end{pmatrix} = -\begin{pmatrix}
G(\mathcal{T}^{\uparrow; e}_{12}) & G(\mathcal{T}^{\uparrow; e}_{13}) & G(\mathcal{T}^{\uparrow; e}_{14}) & S(\mathcal{T}^{\uparrow; e}_{12}) & S(\mathcal{T}^{\uparrow; e}_{13}) & S(\mathcal{T}^{\uparrow; e}_{14})\\
G(\mathcal{T}^{\uparrow; h}_{12}) & G(\mathcal{T}^{\uparrow; h}_{13}) & G(\mathcal{T}^{\uparrow; h}_{14}) & S(\mathcal{T}^{\uparrow; h}_{12}) & S(\mathcal{T}^{\uparrow; h}_{13}) & S(\mathcal{T}^{\uparrow; h}_{14})\\
G(\mathcal{T}^{\uparrow; e}_{22}) & G(\mathcal{T}^{\uparrow; e}_{23}) & G(\mathcal{T}^{\uparrow; e}_{24}) & S(\mathcal{T}^{ \uparrow; e}_{22}) & S(\mathcal{T}^{ \uparrow; e}_{23}) & S(\mathcal{T}^{\uparrow; e}_{24})\\
G(\mathcal{T}^{ \uparrow; h}_{22}) & G(\mathcal{T}^{ \uparrow; h}_{23}) & G(\mathcal{T}^{ \uparrow; h}_{24}) & S(\mathcal{T}^{\uparrow; h}_{22}) & S(\mathcal{T}^{\uparrow; h}_{23}) & S(\mathcal{T}^{\uparrow; h}_{24})\\
G(\mathcal{T}^{\uparrow; e}_{32}) & G(\mathcal{T}^{\uparrow; e}_{33}) & G(\mathcal{T}^{\uparrow; e}_{34}) & S(\mathcal{T}^{\uparrow; e}_{32}) & S(\mathcal{T}^{\uparrow; e}_{33}) & S(\mathcal{T}^{\uparrow; e}_{34})\\
G(\mathcal{T}^{\uparrow; h}_{32}) & G(\mathcal{T}^{\uparrow; h}_{33}) & G(\mathcal{T}^{\uparrow; h}_{34}) & S(\mathcal{T}^{\uparrow; he}_{32}) & S(\mathcal{T}^{\uparrow; h}_{33}) & S(\mathcal{T}^{\uparrow; h}_{34})\\
G(\mathcal{T}^{\uparrow; e}_{42}) & G(\mathcal{T}^{\uparrow; e}_{43}) & G(\mathcal{T}^{\uparrow; e}_{44}) & S(\mathcal{T}^{\uparrow; e}_{42})& S(\mathcal{T}^{\uparrow; e}_{43}) & S(\mathcal{T}^{\uparrow; e}_{44})\\
G(\mathcal{T}^{\uparrow; h}_{42}) & G(\mathcal{T}^{\uparrow; h}_{43}) & G(\mathcal{T}^{\uparrow; h}_{44}) & S(\mathcal{T}^{\uparrow; h}_{42})& S(\mathcal{T}^{\uparrow; h}_{43}) & S(\mathcal{T}^{\uparrow; h}_{44})
\end{pmatrix} \begin{pmatrix}
V\\
V_{3}\\
V_{3}\\
\Delta T_R\\
\Delta T\\
\Delta T
\end{pmatrix}},
\end{equation}}
{The corresponding heat currents due to spin-up electrons and holes are,}
{\begin{equation} \label{eq42}
{\begin{pmatrix}
I_{1q}^{\uparrow, e}\\
I_{1q}^{\uparrow, h}\\
I_{2q}^{\uparrow, e}\\
I_{2q}^{\uparrow, h}\\
I_{3q}^{\uparrow, e}\\
I_{3q}^{\uparrow, h}\\
I_{4q}^{\uparrow, e}\\
I_{4q}^{\uparrow, h}
\end{pmatrix} = -\begin{pmatrix}
TS(\mathcal{T}^{\uparrow; e}_{12}) & TS(\mathcal{T}^{\uparrow; e}_{13}) & TS(\mathcal{T}^{\uparrow; e}_{14}) & L(\mathcal{T}^{\uparrow; e}_{12}) & L(\mathcal{T}^{\uparrow; e}_{13}) & L(\mathcal{T}^{\uparrow; e}_{14})\\
TS(\mathcal{T}^{\uparrow; h}_{12}) & TS(\mathcal{T}^{\uparrow; h}_{13}) & TS(\mathcal{T}^{\uparrow; h}_{14}) & L(\mathcal{T}^{\uparrow; h}_{12}) & L(\mathcal{T}^{\uparrow; h}_{13}) & L(\mathcal{T}^{\uparrow; h}_{14})\\
TS(\mathcal{T}^{\uparrow; e}_{22}) & TS(\mathcal{T}^{\uparrow; e}_{23}) & TS(\mathcal{T}^{\uparrow; e}_{24}) & L(\mathcal{T}^{ \uparrow; e}_{22}) & L(\mathcal{T}^{ \uparrow; e}_{23}) & L(\mathcal{T}^{\uparrow; e}_{24})\\
TS(\mathcal{T}^{ \uparrow; h}_{22}) & TS(\mathcal{T}^{ \uparrow; h}_{23}) & TS(\mathcal{T}^{ \uparrow; h}_{24}) & L(\mathcal{T}^{\uparrow; h}_{22}) & L(\mathcal{T}^{\uparrow; h}_{23}) & L(\mathcal{T}^{\uparrow; h}_{24})\\
TS(\mathcal{T}^{\uparrow; e}_{32}) & TS(\mathcal{T}^{\uparrow; e}_{33}) & TS(\mathcal{T}^{\uparrow; e}_{34}) & L(\mathcal{T}^{\uparrow; e}_{32}) & L(\mathcal{T}^{\uparrow; e}_{33}) & L(\mathcal{T}^{\uparrow; e}_{34})\\
TS(\mathcal{T}^{\uparrow; h}_{32}) & TS(\mathcal{T}^{\uparrow; h}_{33}) & TS(\mathcal{T}^{\uparrow; h}_{34}) & L(\mathcal{T}^{\uparrow; he}_{32}) & L(\mathcal{T}^{\uparrow; h}_{33}) & L(\mathcal{T}^{\uparrow; h}_{34})\\
TS(\mathcal{T}^{\uparrow; e}_{42}) & TS(\mathcal{T}^{\uparrow; e}_{43}) & TS(\mathcal{T}^{\uparrow; e}_{44}) & L(\mathcal{T}^{\uparrow; e}_{42})& L(\mathcal{T}^{\uparrow; e}_{43}) & L(\mathcal{T}^{\uparrow; e}_{44})\\
TS(\mathcal{T}^{\uparrow; h}_{42}) & TS(\mathcal{T}^{\uparrow; h}_{43}) & TS(\mathcal{T}^{\uparrow; h}_{44}) & L(\mathcal{T}^{\uparrow; h}_{42})& L(\mathcal{T}^{\uparrow; h}_{43}) & L(\mathcal{T}^{\uparrow; h}_{44})
\end{pmatrix} \begin{pmatrix}
V\\
V_{3}\\
V_{3}\\
\Delta T_R\\
\Delta T\\
\Delta T
\end{pmatrix}},
\end{equation}}
{Similarly, the total {charge} current due to spin-down electrons and holes is given by,}

{\begin{equation} \label{eq43}
{\begin{pmatrix}
I_{1c}^{\downarrow, e}\\
I_{1c}^{\downarrow, h}\\
I_{2c}^{\downarrow, e}\\
I_{2c}^{\downarrow, h}\\
I_{3c}^{\downarrow, e}\\
I_{3c}^{\downarrow, h}\\
I_{4c}^{\downarrow, e}\\
I_{4c}^{\downarrow, h}
\end{pmatrix} = -\begin{pmatrix}
G(\mathcal{T}^{\downarrow; e}_{12}) & G(\mathcal{T}^{\downarrow; e}_{13}) & G(\mathcal{T}^{\downarrow; e}_{14}) & S(\mathcal{T}^{\downarrow; e}_{12}) & S(\mathcal{T}^{\downarrow; e}_{13}) & S(\mathcal{T}^{\downarrow; e}_{14})\\
G(\mathcal{T}^{\downarrow; h}_{12}) & G(\mathcal{T}^{\downarrow; h}_{13}) & G(\mathcal{T}^{\downarrow; h}_{14}) & S(\mathcal{T}^{\downarrow; h}_{12}) & S(\mathcal{T}^{\downarrow; h}_{13}) & S(\mathcal{T}^{\downarrow; h}_{14})\\
G(\mathcal{T}^{\downarrow; e}_{22}) & G(\mathcal{T}^{\downarrow; e}_{23}) & G(\mathcal{T}^{\downarrow; e}_{24}) & S(\mathcal{T}^{ \downarrow; e}_{22}) & S(\mathcal{T}^{ \downarrow; e}_{23}) & S(\mathcal{T}^{\downarrow; e}_{24})\\
G(\mathcal{T}^{ \downarrow; h}_{22}) & G(\mathcal{T}^{ \downarrow; h}_{23}) & G(\mathcal{T}^{ \downarrow; h}_{24}) & S(\mathcal{T}^{\downarrow; h}_{22}) & S(\mathcal{T}^{\downarrow; h}_{23}) & S(\mathcal{T}^{\downarrow; h}_{24})\\
G(\mathcal{T}^{\downarrow; e}_{32}) & G(\mathcal{T}^{\downarrow; e}_{33}) & G(\mathcal{T}^{\downarrow; e}_{34}) & S(\mathcal{T}^{\downarrow; e}_{32}) & S(\mathcal{T}^{\downarrow; e}_{33}) & S(\mathcal{T}^{\downarrow; e}_{34})\\
G(\mathcal{T}^{\downarrow; h}_{32}) & G(\mathcal{T}^{\downarrow; h}_{33}) & G(\mathcal{T}^{\downarrow; h}_{34}) & S(\mathcal{T}^{\downarrow; he}_{32}) & S(\mathcal{T}^{\downarrow; h}_{33}) & S(\mathcal{T}^{\downarrow; h}_{34})\\
G(\mathcal{T}^{\downarrow; e}_{42}) & G(\mathcal{T}^{\downarrow; e}_{43}) & G(\mathcal{T}^{\downarrow; e}_{44}) & S(\mathcal{T}^{\downarrow; e}_{42})& S(\mathcal{T}^{\downarrow; e}_{43}) & S(\mathcal{T}^{\downarrow; e}_{44})\\
G(\mathcal{T}^{\downarrow; h}_{42}) & G(\mathcal{T}^{\downarrow; h}_{43}) & G(\mathcal{T}^{\downarrow; h}_{44}) & S(\mathcal{T}^{\downarrow; h}_{42})& S(\mathcal{T}^{\downarrow; h}_{43}) & S(\mathcal{T}^{\downarrow; h}_{44})
\end{pmatrix} \begin{pmatrix}
V\\
V_{3}\\
V_{3}\\
\Delta T_R\\
\Delta T\\
\Delta T
\end{pmatrix}},
\end{equation}}
{and the corresponding heat currents due to spin-down electrons and holes as,}
{\begin{equation} \label{eq44}
{\begin{pmatrix}
I_{1q}^{\downarrow, e}\\
I_{1q}^{\downarrow, h}\\
I_{2q}^{\downarrow, e}\\
I_{2q}^{\downarrow, h}\\
I_{3q}^{\downarrow, e}\\
I_{3q}^{\downarrow, h}\\
I_{4q}^{\downarrow, e}\\
I_{4q}^{\downarrow, h}
\end{pmatrix} = -\begin{pmatrix}
TS(\mathcal{T}^{\downarrow; e}_{12}) & TS(\mathcal{T}^{\downarrow; e}_{13}) & TS(\mathcal{T}^{\downarrow; e}_{14}) & L(\mathcal{T}^{\downarrow; e}_{12}) & L(\mathcal{T}^{\downarrow; e}_{13}) & L(\mathcal{T}^{\downarrow; e}_{14})\\
TS(\mathcal{T}^{\downarrow; h}_{12}) & TS(\mathcal{T}^{\downarrow; h}_{13}) & TS(\mathcal{T}^{\downarrow; h}_{14}) & L(\mathcal{T}^{\downarrow; h}_{12}) & L(\mathcal{T}^{\downarrow; h}_{13}) & L(\mathcal{T}^{\downarrow; h}_{14})\\
TS(\mathcal{T}^{\downarrow; e}_{22}) & TS(\mathcal{T}^{\downarrow; e}_{23}) & TS(\mathcal{T}^{\downarrow; e}_{24}) & L(\mathcal{T}^{ \downarrow; e}_{22}) & L(\mathcal{T}^{ \downarrow; e}_{23}) & L(\mathcal{T}^{\downarrow; e}_{24})\\
TS(\mathcal{T}^{ \downarrow; h}_{22}) & TS(\mathcal{T}^{ \downarrow; h}_{23}) & TS(\mathcal{T}^{ \downarrow; h}_{24}) & L(\mathcal{T}^{\downarrow; h}_{22}) & L(\mathcal{T}^{\downarrow; h}_{23}) & L(\mathcal{T}^{\downarrow; h}_{24})\\
TS(\mathcal{T}^{\downarrow; e}_{32}) & TS(\mathcal{T}^{\downarrow; e}_{33}) & TS(\mathcal{T}^{\downarrow; e}_{34}) & L(\mathcal{T}^{\downarrow; e}_{32}) & L(\mathcal{T}^{\downarrow; e}_{33}) & L(\mathcal{T}^{\downarrow; e}_{34})\\
TS(\mathcal{T}^{\downarrow; h}_{32}) & TS(\mathcal{T}^{\downarrow; h}_{33}) & TS(\mathcal{T}^{\downarrow; h}_{34}) & L(\mathcal{T}^{\downarrow; he}_{32}) & L(\mathcal{T}^{\downarrow; h}_{33}) & L(\mathcal{T}^{\downarrow; h}_{34})\\
TS(\mathcal{T}^{\downarrow; e}_{42}) & TS(\mathcal{T}^{\downarrow; e}_{43}) & TS(\mathcal{T}^{\downarrow; e}_{44}) & L(\mathcal{T}^{\downarrow; e}_{42})& L(\mathcal{T}^{\downarrow; e}_{43}) & L(\mathcal{T}^{\downarrow; e}_{44})\\
TS(\mathcal{T}^{\downarrow; h}_{42}) & TS(\mathcal{T}^{\downarrow; h}_{43}) & TS(\mathcal{T}^{\downarrow; h}_{44}) & L(\mathcal{T}^{\downarrow; h}_{42})& L(\mathcal{T}^{\downarrow; h}_{43}) & L(\mathcal{T}^{\downarrow; h}_{44})
\end{pmatrix} \begin{pmatrix}
V\\
V_{3}\\
V_{3}\\
\Delta T_R\\
\Delta T\\
\Delta T
\end{pmatrix}},
\end{equation}}
\end{widetext}}
{where $G(X)$ is the electrical conductance for particles with transmission probability $X$, $S(X)$ is the Seebeck coefficient for particles with transmission probability $X$, and $L(X)$ is the thermal conductance for particles with transmission probability $X$. The electrical, Seebeck, and thermal conductances are given in Eqs. (\ref{eq22}-\ref{eq24}).}
\bibliography{apssamp5}
\end{document}